\documentclass[
 superscriptaddress,
 reprint,
 amsmath, amssymb,
 aps, preprintnumbers, nofootinbib, floatfix, longbibliography
]{revtex4-1}

\usepackage{graphicx}
\usepackage{dcolumn}
\usepackage{bm}
\usepackage{color}
\usepackage{amssymb}
\usepackage{amsmath}
\usepackage[colorlinks=true,allcolors=Purple,pdfusetitle]{hyperref}
\usepackage{rotating}
\usepackage{multirow}
\usepackage[dvipsnames]{xcolor}
\usepackage{svg}
\usepackage{esdiff}
\usepackage{comment}
\usepackage[T1]{fontenc}
\definecolor{Green}{RGB}{0, 128, 0}
\newcommand{\orcid}[1]{\href{https://orcid.org/#1}{\includegraphics[width=10pt]{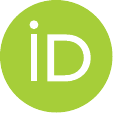}}}
\newcommand{\SN}{\vec{S}_N}
\newcommand{\Schi}{\vec{S}_\chi}
\newcommand{\qmN}{\frac{\vec{q}}{m_N}}

\begin{document}
\preprint{N3AS-23-011}
\preprint{INT-PUB-23-016}

\title{Uncertainties on the EFT coupling limits for direct dark matter detection experiments stemming from uncertainties of target properties}

\author{Daniel~J.~Heimsoth~\orcid{0000-0002-5110-1704}}
\email{dheimsoth@wisc.edu}
\affiliation{
Department of Physics, University of Wisconsin--Madison,
Madison, Wisconsin 53706, USA}

\author{Brandon~Lem~\orcid{0000-0001-6708-7452}}
\email{brandonlem@berkeley.edu}
\affiliation{
Department of Physics, University of California Berkeley, Berkeley, California 94720, USA}

\author{Anna~M.~Suliga~\orcid{0000-0002-8354-012X}}
\email{asuliga@berkeley.edu}
\affiliation{
Department of Physics, University of California Berkeley, Berkeley, California 94720, USA}
\affiliation{
Department of Physics, University of Wisconsin--Madison,
Madison, Wisconsin 53706, USA}

\author{Calvin~W.~Johnson~\orcid{0000-0003-1059-7384}}
\email{cjohnson@sdsu.edu}
\affiliation{
Department of Physics, San Diego State University, San Diego, California 92182-1233, USA}

\author{A.~Baha~Balantekin~\orcid{0000-0002-2999-0111}}
\email{baha@physics.wisc.edu}
\affiliation{
Department of Physics, University of Wisconsin--Madison,
Madison, Wisconsin 53706, USA}

\author{Susan~N.~Coppersmith~\orcid{0000-0001-6181-9210}}
\email{snc@physics.wisc.edu}
\affiliation{
Department of Physics, University of Wisconsin--Madison,
Madison, Wisconsin 53706, USA}
\affiliation{
School of Physics, The University of New South Wales, Sydney, New South Wales 2052, Australia}

\date{October 19, 2023}


\begin{abstract}

Direct detection experiments are still one of the most promising ways to unravel the nature of dark matter. To fully understand how well these experiments constrain the dark matter interactions with the Standard Model particles, all the uncertainties affecting the calculations must be known. It is especially critical now because direct detection experiments recently moved from placing limits only on the two elementary spin independent and spin dependent operators to the complete set of possible operators coupling dark matter and nuclei in nonrelativistic theory. In our work, we estimate the effect of nuclear configuration-interaction uncertainties on the exclusion bounds for one of the existing xenon-based experiments for all fifteen operators. We find that for operator number 13 the $\pm1\sigma$ uncertainty on the coupling between the dark matter and nucleon can reach more than 50\% for dark matter masses between 10 and 1000 GeV. In addition, we discuss how quantum computers can help to reduce this uncertainty and how the uncertainties are affected for couplings obtained for the nonrelativistic reductions of the relativistic interactions.

\end{abstract}

\maketitle


\section{Introduction}
\label{sec:Introduction}

The nature of dark matter (DM) has remained one of the biggest mysteries in physics. Over the past few decades, experiments have attempted to measure DM interactions with visible matter, both directly and indirectly, to no avail. Direct detection experiments for particle DM~(see, e.g., Ref.~\cite{Akerib:2022ort} for a recent review), among which the currently largest and leading ones are XENONnT~\cite{XENONCollaboration:2022kmb, XENON:2023sxq} and PandaX-4T~\cite{PandaX-4T:2021bab}, are generally built to search an area of parameter space consistent with a particular class of DM models, such as the Weakly Interacting Massive Particle (WIMP), see, e.g., Ref.~\cite{Roszkowski:2017nbc} for a recent review. These experiments frequently look at only the leading order spin-independent and spin-dependent interactions~\cite{LUX:2017ree, XENON:2018voc, DarkSide:2018bpj, LZ:2022ufs}, which may be suppressed by an as-yet-unknown mechanism that prevents any detection of DM. Recently, there has been heightened interest to focus on a more general approach in the Galilean Effective Field Theory (EFT)~\cite{Fitzpatrick:2012ix, Anand:2013yka} framework for setting constraints on DM-nuclei coupling and cross sections~\cite{DarkSide-50:2020swd, LUX:2021ksq} for elastic scattering, as well as in Chiral Effective Field Theory for inelastic scattering~\cite{XENON:2022avm}. 
{\color{black} We note that while here we follow  effective theories developed at the level of nucleon 
fields~\cite{Fitzpatrick:2012ix, Anand:2013yka}, other effective theories start at the level of quark and gluon fields~\cite{Cirigliano:2012pq,hoferichter2015chiral, Bishara_2017,Brod:2017bsw,hoferichter2019darkmatternucleus,PhysRevD.99.055031,PhysRevC.106.044003,XENON:2022avm,Trickle:2020oki}.
}

Besides detectors built for the main goal of DM detection, other low-energy, rare event search experiments can also be used in DM direct searches. Neutrinoless double-beta decay ($0\nu\beta\beta$) is currently an extremely rare interaction of interest, and detectors such as CUORE designed for $0\nu\beta\beta$ also look for DM scattering events off their target material~\cite{ADAMS2022103902}. CUORE uses TeO$_2$ crystals as a target; tellurium, like xenon (which we focus our analysis on in this paper), has isotopes that are currently  modeled through  nuclear {\color{black} configuration-interaction approaches} because of the large number of nucleons ($\sim$~120-140). Thus, our calculations of WIMP-nucleon coupling uncertainties will be applicable to a wide range of sensitive, low-energy experiments.

In this paper, we show that a major source of uncertainty in the calculation of Galilean EFT DM-nucleus cross sections and the upper limits on DM couplings is nuclear models. Nuclei are highly complex many-particle quantum systems and thus practically impossible to model or simulate to high accuracy without simplifying approximations on modern hardware. 
A common approach for heavier nuclei is the nuclear {\color{black} configuration-interaction in a} shell model basis, where a frozen core of nucleons is surrounded by valence nucleons which occupy energy and angular momentum states in close analogy with atomic electrons~\cite{Suhonen2007}. 
{\color{black} (\textit{Ab initio} calculations of nuclei have been developed over recent years \cite{Hergert:2020bxy}, but only very recently have these techniques started to be applied to heavy, open-shell nuclides such as xenon isotopes~\cite{PhysRevLett.128.072502}.)}
The {\color{black} target} nuclear response to {\color{black}external probes such as DM scattering} is then encoded in reduced matrix elements and a density matrix which contains the transition probabilities between different states of valence nucleon configurations. Still, this is a many-body system and thus challenging to calculate numerically. 

\begin{figure*}
    \centering
    \includegraphics[width=1.5\columnwidth]{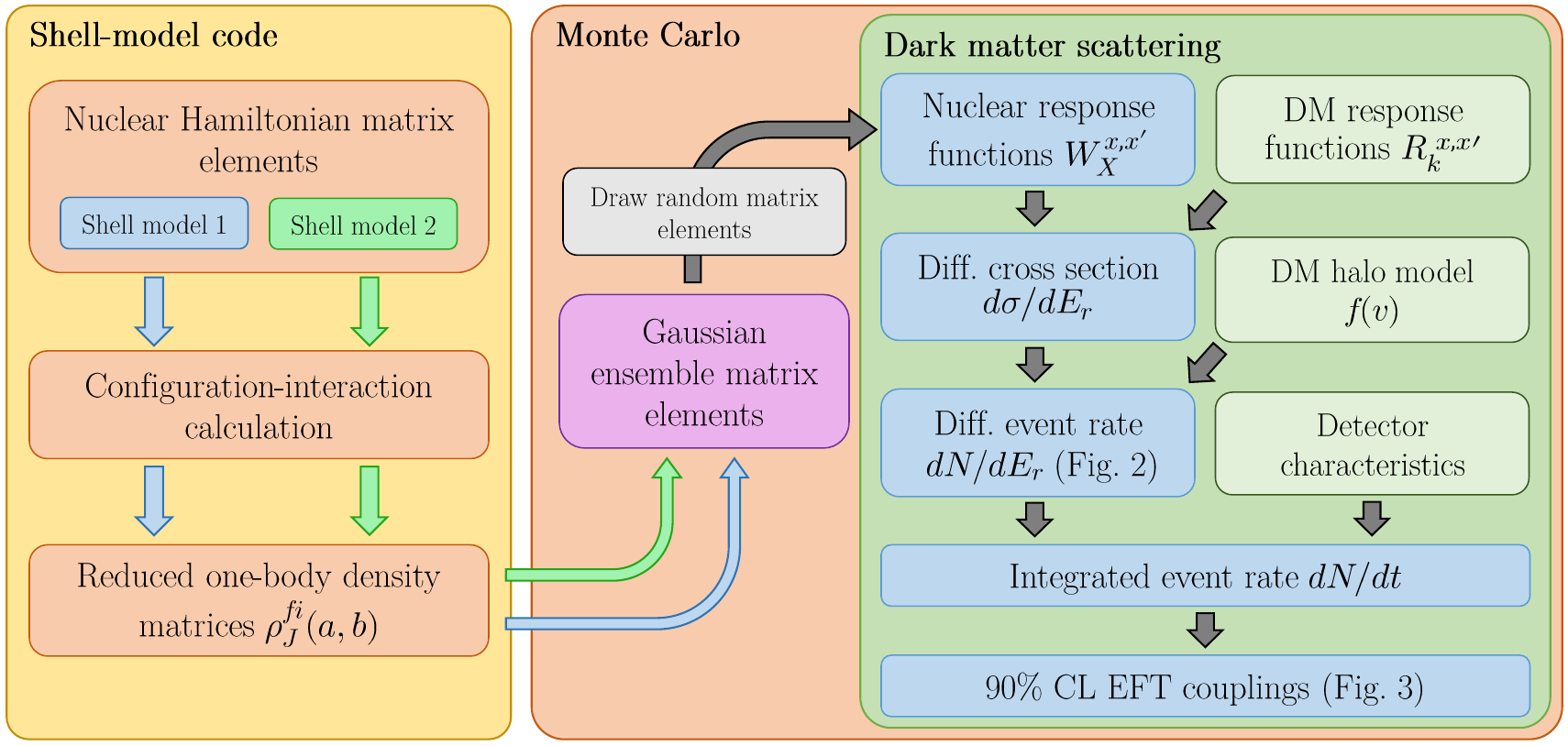}
    \caption{A schematic describing the general process to calculate upper limits on DM-nucleon EFT coupling coefficients and their uncertainties propagated from uncertainties in {\color{black} configuration-interaction calculations} for a particular isotope. In our work, the two {\color{black}configuration-interaction Hamiltonians} are GCN and JJ55 (described in Sec.~\ref{sec:Nuclear-response-functions}). 
    Our Monte Carlo procedure to construct Gaussian distributions for each one-body density matrix element ($\rho_J^{fi} (a,b)$, see Eq.~\eqref{eq:one-body-density-matrix}) is described in detail in Sec.~\ref{sec:Tools-and-Methods}. The differential DM scattering rates ($dN/dE_\mathrm{r}$, with $N$ the event rate and $E_r$ the recoil energy, see Eq.~\eqref{eq:diffeventrate}) are calculated using the \emph{dmscatter} code which implements the intermediate calculational steps such as nuclear response function ($W_X^{x,x^\prime}$, see Eq.~\eqref{eq:W_xx}), differential cross sections ($d\sigma/dE_\mathrm{r}$, see Eq.~\eqref{eq:cross-section}), DM halo model ($f(v)$, see Eq.~\eqref{eq:DM-density-distribiution}), and DM response function $R_k^{x,x^\prime}$ (see, eg., App.~C of Ref.~\cite{Gorton:2022eed}, or Eq.~38 in Ref.~\cite{Fitzpatrick:2012ix}). Then the energy integrated rates $dN/dt$ were calculated adding the realistic XENON1T energy window, efficiency, and exposure to obtain the upper limits on the DM EFT coupling coefficients and their uncertainties as decribed in Sec.~\ref{sec:detector-limits}.}
    \label{fig:flowchart}
\end{figure*}

In our work, we quantify the uncertainty in the upper limits of the Galilean EFT DM-nucleus couplings by considering the uncertainty in {\color{black}configuration-interaction calculations of the target response}. Figure~\ref{fig:flowchart} illustrates our process {\color{black} and provides a map of this paper}. First, we choose as initial inputs two different appropriate {\color{black}configuration-interaction Hamiltonians, from which one generates the respective ground states and subsequently} the relevant {\color{black}target} one-body density matrices. From those computed density matrices we define a simple ensemble of density matrices, 
 use a Monte Carlo (MC) procedure to sample matrices randomly from that ensemble, 
 and characterize the resulting spread in the dependence of scattering event rates on recoil energy. Finally, we calculate the upper limits on the Galilean EFT WIMP-nucleus couplings and their uncertainties coming from differences between the two model {\color{black}Hamiltonians}. {\color{black} As discussed in Section~\ref{sec:Tools-and-Methods}, this is a simplified approach to uncertainty quantification.}

We find that the uncertainties in the upper limits of the coupling coefficients vary significantly between the different operators.  The uncertainties for upper limits on couplings to some of the operators can be substantial, up to an order of magnitude for the $\pm1\sigma$ region.
These results suggest that nuclear uncertainties need to be taken into account by experiments when placing upper limits on the DM-nucleus coupling coefficients for some operators.  In addition, improvements in calculations of nuclear structure could significantly improve the accuracy of results emerging from dark matter detection experiments.

The remainder of this paper is organized as follows.
Sec.~\ref{sec:theory} presents the theoretical background used to calculate differential event rates for the 15 different possible nonrelativistic interaction channels between DM and nuclei.
In Sec.~\ref{sec:Tools-and-Methods} we present our methods to model the uncertainties coming from {\color{black}configuration-interaction} calculations of the DM scattering rates.
Sec.~\ref{sec:results} shows the results of the calculations for the ideal differential event rates and their uncertainties for different interaction channels.
Then, in Sec.~\ref{sec:detector-limits} we use our results together with knowledge of the properties of the XENONIT experiment to determine upper limits on the WIMP-nucleon coupling together with the propagated shell model uncertainties. 
In Sec.~\ref{sec:Discussion} we compare the uncertainties arising from limits to our knowledge of the nuclear models to other sources of uncertainty. 
In addition, we show how the uncertainties on the coupling limits vary for the operators coming from nonrelativistic reductions of the relativistic interactions. We also discuss ways to improve the estimates of nuclear uncertainties as well as methods for reducing these uncertainties, including the possible utility of quantum computation in this context. 
In Sec.~\ref{sec:Conclusions} we conclude. 

\section{Theoretical background: Nonrelativistic EFT Operators Describing Possible DM-Nuclei Interactions}
\label{sec:theory}

This section outlines the theory behind the  differential WIMP-nucleus event rates.
 First, in Sec.~\ref{sec:Operators} we briefly review the nonrelativistic theory for DM-nuclei interactions, listing the derived operators for DM-nuclei elastic scattering. Next, in Sec.~\ref{sec:Nuclear-response-functions} we show how the nuclear response functions are built from the nuclear density matrix elements, while in Sec.~\ref{sec:Event-rates} we present expressions for the differential cross-sections as a function of energy when the relevant nuclear response functions are known precisely.

\subsection{Nonrelativistic Operators}
\label{sec:Operators}

The EFT calculations developed in Refs.~\cite{Fitzpatrick:2012ix, Anand:2013yka}, which assume Galilean invariance, lead to fifteen operators that are at most second-order in the exchanged momentum operators and are built as products of the following quantities:
\begin{equation*}
\label{eq:building-blocks}
    i \vec{q}, \quad \vec{v}^{\perp} \equiv \vec{v} + \frac{\vec{q}}{2 \mu_N}, \quad \vec{S}_\chi, \quad \vec{S}_N,
\end{equation*}
where $\mu_N = \frac{m_\chi m_N}{m_\chi + m_N}$ is the dark matter-nucleon reduced mass with $m_\chi$ ($m_N$) being the WIMP (nucleon) mass, $\Schi$ is the WIMP spin and $\SN$ is the nucleon spin, $\vec{q}$ is the exchanged momentum, $\vec{v}$ is the relative incoming velocity, and $\vec{v}^\perp \cdot \vec{q} = 0$ by momentum conservation.

These nonrelativistic EFT operators are~\cite{Fitzpatrick:2012ix, Anand:2013yka}:
\begin{subequations}
\label{eq:operators} 
\begin{eqnarray}
	\mathcal{O}_1 &=& 1_\chi 1_N \ , \label{eq:O1} \\
	\mathcal{O}_2 &=& (v^\perp)^2 \ , \label{eq:O2} \\
	\mathcal{O}_3 &=& i \SN \cdot \left( \qmN \times \vec{v}^\perp \right) \ , \label{eq:O3} \\
	\mathcal{O}_4 &=& \SN \cdot \Schi \ , \label{eq:O4}\\
	\mathcal{O}_5 &=& i \Schi \cdot \left( \qmN \times \vec{v}^\perp \right) \ , \label{eq:O5} \\
	\mathcal{O}_6 &=& \left( \Schi \cdot \qmN \right) \left( \SN \cdot \qmN \right)  \ , \label{eq:O6} \\
	\mathcal{O}_7 &=& \SN \cdot \vec{v}^\perp \ , \label{eq:O7} \\
	\mathcal{O}_8 &=& \Schi \cdot \vec{v}^\perp \ , \label{eq:O8} \\
	\mathcal{O}_9 &=& i \Schi \cdot \left( \SN \times \qmN \right) \ , \label{eq:O9} \\
	\mathcal{O}_{10} &=& i\SN \cdot \qmN \ , \label{eq:O10} \\
	\mathcal{O}_{11} &=& i\Schi \cdot \qmN \ , \label{eq:O11} \\
	\mathcal{O}_{12} &=& \Schi \cdot \left( \SN \times \vec{v}^\perp \right) \ , \label{eq:O12} \\
	\mathcal{O}_{13} &=& i \left( \Schi \cdot \vec{v}^\perp \right) \left( \SN \cdot \qmN \right) \ , \label{eq:O13} \\
	\mathcal{O}_{14} &=& i \left( \Schi \cdot \qmN \right) \left( \SN \cdot \vec{v}^\perp \right) \ , \label{eq:O14} \\
	\mathcal{O}_{15} &=& - \left( \Schi \cdot \qmN \right) \left[ \left( \SN \times \vec{v}^\perp \right) \cdot \qmN \right] \ , \label{eq:O15}
\end{eqnarray}
\end{subequations}
where $N = \text{n,p}$ for neutrons and protons, respectively. The total interaction Hamiltonian including the responses from all the operators is given by
\begin{equation}
\label{eq:Total-Hamiltonian}
	\mathcal{H} = \sum_{x=\text{n,p}} \sum_{i=1}^{15} c_i^{x} \mathcal{O}_i^{x} \ ,
\end{equation}
where the EFT coupling coefficients $c_{i}^{x}$, with $i=1,..,15$ being the operator index and $x$ the nucleon type, are \textit{a priori} unknown. In Sec.~\ref{sec:detector-limits} we show how their values can be constrained by experimental non-detection of WIMPs with the data from the existing DM direct detectors.

\subsection{Target response functions}
\label{sec:Nuclear-response-functions}

The probability of a DM-nucleus interaction, i.e., the cross section, can be factorized into the dark matter response functions $R_{k}^{x,x'}$ and the {\color{black}target} response functions $W_{k}^{x,x'}$, where the index $k = 1,..,8$ denotes the allowed combinations of electro-weak-theory operators. 

The DM response functions, which group the operators derived from Galilean EFT theory (Eq.~\eqref{eq:operators}) for the corresponding nuclear response functions,  
can be found in App.~C of Ref.~\cite{Gorton:2022eed}, or Eq.~38 in Ref.~\cite{Fitzpatrick:2012ix}.

The {\color{black}target} response functions are built from eight one-body electroweak multipole operators $X_{J_T}$, 
\begin{equation}
\label{eq:W_xx}
    W_X^{x,x'} = \sum_{{J_T}} \langle \Psi_f || X_{J_T}^x || \Psi_i \rangle \langle \Psi_i || X_{J_T}^{x'} || \Psi_f \rangle \ ,
\end{equation}
where $\Psi_i$ is the initial target wave function  and $\Psi_f$ is the final target wave function, {\color{black} and $J_T$ is 
the angular momentum rank of the operator}. 
The matrix elements of these operators,
which  can be constructed from Bessel spherical harmonics and vector harmonics~\cite{Donnelly:1979ezn},
are calculated by summing over single particle orbitals from a one-body density matrix $\rho$:
\begin{equation}
\label{eq:density-matrix}
    \langle \Psi_f || X_{J_T}^x || \Psi_i \rangle = \sum_{a,b} \langle a || X_{J_T}^x || b \rangle \rho_J^{fi} (a,b) \ ,
\end{equation}
where $a$ and $b$ are the indices of the single-particle orbitals, and $\langle a || X_{J_T}^x || b \rangle$ are the reduced matrix elements of the operator $X_{J_T}^x$, as obtained from the Wigner-Eckart theorem~\cite{edmonds1957angular}. 

The target one-body density matrices are
\begin{equation}
\label{eq:one-body-density-matrix}
    \rho_{J_T}^{fi}(a,b) = \frac{1}{\sqrt{2{J_T} + 1}} \langle \Psi_f || [\hat{c}_a^{\dagger} \otimes \Tilde{c}_b]_{J_T} || \Psi_i \rangle \quad ,
\end{equation}
where $\hat{c}_a^{\dagger}$ and $\Tilde{c}_b$ are the fermion creation and destruction operators~\cite{edmonds1957angular}.
{\color{black} The wave functions and the subsequent density matrices used in this work were computed~\cite{Gorton:2022eed}
using in configuration-interaction in a shell model basis 
via a high-performance code~\cite{BIGSTICK,Johnson:2018hrx}.}

Since we assume the WIMP mass to be at least on the order of 10-1000 GeV and the WIMPs to be nonrelativistic, relatively little energy ($\mathcal{O}(100 ~\text{keV})$) is transferred to the nucleus. As such, we expect that the nucleus stays in the ground state throughout the interaction, i.e., $| \Psi_i \rangle  = | \Psi_f \rangle $. Thus, we are only interested in ground state-to-ground state transitions.

In our work, {\color{black} we use eigenstates and the resulting density matrices computed in a fixed Hilbert space  using two sets of nuclear Hamiltonian matrix elements.  The Hilbert space has a fixed core of $^{100}$Sn and a valence space comprised of the  $0g_{7/2}$-$2s_{1/2}$-$1d_{3/2}$-$1d_{5/2}$-$0h_{11/2}$ orbitals. The two Hamiltonians are labeled as}  GCN~\cite{Caurier:2007wq, Caurier:2010az} and JJ55~\cite{Brown:2004xk}. The GCN shell model interaction was fit to experimental nuclear energies of nuclides in this mass region, {\color{black} starting from} a $G$-matrix~\cite{HJORTHJENSEN1995125} constructed using the charge-dependent Bonn  nucleon-nucleon (NN) potential version C, also known as the CD-Bonn potential~\cite{Machleidt:2000ge}. 
The JJ55 interaction is also obtained starting from a CD-Bonn $G$-matrix but was fitted to different experimental data from this mass region. 
Thus, {\color{black} the differences in the calculated target  states} represent the challenges in accurately modeling the {\color{black} nuclear many-body system.}

\subsection{Dark matter-nucleus event rates}
\label{sec:Event-rates}

The differential event rate $\text{d}N/\text{d}E_\text{r}$ ($N$ is the number of events and $E_\mathrm{r}$ is the recoil energy) is the convolution of the differential cross section $\mathrm{d}\sigma/\mathrm{d}E_\text{r}$ for the WIMP-nucleus interaction and a DM halo model, integrated over DM velocity $v$:
\begin{equation}\label{eq:diffeventrate}
    \frac{\mathrm{d}N}{\mathrm{d}E_\text{r}} = \varepsilon (E_r) N_T n_\chi \int \frac{\mathrm{d}\sigma}{\mathrm{d}E_\text{r}} \tilde{f}(\vec{v}) |\vec{v}| \ \mathrm{d}^3\vec{v} \quad .
\end{equation}
In the above equation, $\varepsilon (E_r)$ is the detector efficiency, $N_T$ is the number of target nuclei in the detector, $n_\chi$ is the local DM number density, and $\tilde{f}(\vec{v})$, the DM velocity distribution in the rest frame of the detector, is obtained by boosting from the DM velocity distribution as seen in the rest frame of the Sun (the galactic-frame distribution). For this galactic-frame distribution, we use the simple halo model given by~\cite{Drukier:1986tm, PhysRevD.37.3388}
\begin{equation}
\label{eq:DM-density-distribiution}
    f(\vec{v}) = \frac{\Theta (v_{\text{esc}} - |\vec{v}|)}{\pi^{3/2} v_0^3 N_{\text{esc}}} e^{-\vec{v}^2 / v_0^2} \ ,
\end{equation}
where $\Theta(v)$ is the Heaviside step function, $v_0 = 220$~km/s, $N_{\text{esc}}$ is a normalization factor, and the DM velocity cutoff $v_{\text{esc}} = 550$~km/s, roughly the galactic escape velocity.

The differential cross section, $\mathrm{d}\sigma/\mathrm{d}E_\text{r}$ in Eq.~\eqref{eq:diffeventrate}, for a DM-nucleus interaction for a particular isotope  can be calculated as follows~\cite{Gorton:2022eed}
\begin{equation}
\label{eq:cross-section}
    \frac{\mathrm{d}\sigma}{\mathrm{d}E_\text{r}} = \frac{2m_T}{4\pi v^2} T(v,q) \ ,
\end{equation}
where $m_T$ is the mass of the target nucleus (i.e., $m_T = A m_N$ where $A$ is the mass number of the nucleus and $m_N = 0.938$~GeV/$c^2$ is the nucleon mass) and $q = \sqrt{2m_TE_r}$ is the momentum transfer. The scattering transition probability~\cite{Fitzpatrick:2012ix, Anand:2013yka, Gorton:2022eed} is given by
\begin{equation}
\label{eq:transition-probability}
    T(v,q) = \frac{4\pi}{2j_T+1} \sum_{x=\mathrm{p,n}} \sum_{x'=\mathrm{p,n}} \sum_{i=1}^8 R_i^{x,x'} W_i^{x,x'} \ ,
\end{equation}
where $j_T$ is the spin of the target nucleus and the sums are factored out into two parts, as mentioned in Sec.~\ref{sec:Nuclear-response-functions}. The first part contains the particle physics, i.e., the WIMP response function $R_i^{x,x'}$, and the second part contains the nuclear physics, i.e., the {\color{black}target} response functions $W_X^{x,x'}$ given in Eq.~\eqref{eq:W_xx}.

\section{Methods: Modeling uncertainties in dark matter - target event rates}
\label{sec:Tools-and-Methods}
In this section, we introduce our strategy for calculating the impact of uncertainty from the nuclear shell models on the DM--nucleus event rates.

{\color{black}  To carry out uncertainty quantification (UQ) 
of DM detection is a daunting problem. 
A full UQ effort might involve hundreds or thousands of Monte Carlo (MC) samples of the input Hamiltonian matrix elements~\cite{PhysRevC.98.061301,fox_usdb,fox2022uncertainty}, the first box of Figure~\ref{fig:flowchart}. The large configuration-interaction dimensions of the target isotopes, up to tens of billions, 
require thousands of hours of supercomputer time for each Hamiltonian sampled, an impractical methodology.  Instead we choose to sample a model of the one-body density 
matrices, a procedure which may not 
fully reflect the uncertainties of the 
underlying Hamiltonian. Future work might use emulators, an increasingly popular approach in UQ~\cite{konig_ec}, to speed up the sampling of the Hamiltonian, but emulators have yet to be fully applied to  configuration-interaction 
calculations.}

To estimate the impact of the uncertainties coming from {\color{black}configuration-interaction} calculations on the upper limits of the WIMP-nucleon coupling, we use a simple MC procedure.
As inputs to our MC simulations, we use two previously calculated nuclear structure data files containing reduced one-body density matrices calculated in the configuration-interaction shell model~\cite{Gorton:2022eed,BIGSTICK,Johnson:2018hrx} for two different {\color{black}configuration-interaction Hamiltonians}, GCN~\cite{Caurier:2007wq, Caurier:2010az}, and JJ55~\cite{Brown:2004xk}, which are described in more detail in Sec.~\ref{sec:Nuclear-response-functions}.
This is represented in the flowchart in 
Fig.~\ref{fig:flowchart} by the  box ``Shell-model code.''

{\color{black} To provide an approximate model of an ensemble of calculations,}
we calculate the mean and standard deviation of each density matrix element $\rho_{J_T}^{fi}(a,b)$ between the two {\color{black} interactions}, i.e., for all combinations of initial and final orbitals $a$ and $b$.  With these averages and standard deviations, we construct a Gaussian distribution for each of the density matrix elements.  Then, we randomly draw a value for each element from the defined distributions {\color{black} (This is in lieu of a MC sampling over an ensemble of Hamiltonians, which would be more fundamental but, as stated above, is not currently computationally tractable.)} There is one caveat: the ground state-to-ground state density matrix elements for the $J_T = 0$ transition for both protons and neutrons are each constrained by a sum rule in order to conserve particle number:
\begin{subequations}
\label{eq:sum-rules} 
\begin{eqnarray}
	\sum_a \rho_{\text{p},J_T=0}(a,a) \frac{[j_a]}{[J_i]} = Z_\mathrm{valence} \ , \label{eq:sum-protons} \\
	\sum_a \rho_{\text{n},J_T=0}(a,a) \frac{[j_a]}{[J_i]} = N_\mathrm{valence} \ , \label{eq:sum-neutrons}
\end{eqnarray}
\end{subequations}
where $j_a$ is the total angular momentum of the $a$-th valence orbital, $J_i$ is the total angular momentum of 
{\color{black} the state $\Psi$ in Eq.~(\ref{eq:one-body-density-matrix}), in our case always 
the ground state;} $Z_\mathrm{valence}$ and $N_\mathrm{valence}$ are the number of valence protons and neutrons, respectively, and the bracket notation is $[x] \equiv \sqrt{2x+1}$.  We normalize the randomly generated $J_T=0$ density matrix element values such that the sum rules in Eqs.~\eqref{eq:sum-protons} and~\eqref{eq:sum-neutrons} are satisfied. {\color{black} There are  no analogous sum rules for $J_T > 0$, so it is possible to have unphysical values for those density matrices. Since our ensemble is defined by physical values, however, we nonetheless assume our calculations are 
sufficient to provide a realistic estimate of uncertainties.}

The procedure described above can be briefly summarized with these three steps:
\begin{enumerate}
    \item{Construct a Gaussian distribution for each one-body density matrix element $\rho_{J_T}^{fi}(a,b)$ using matrices generated with at least two different nuclear shell models,}
    \item{Draw randomized nuclear density matrix elements from created Gaussian distributions,}
    \item{Normalize the new elements to satisfy particle number conservation, i.e., Eqs.~\eqref{eq:sum-protons} and~\eqref{eq:sum-neutrons}.}
\end{enumerate}

The ``Monte Carlo'' section of the flowchart in Fig.~\ref{fig:flowchart} depicts the MC process we have just described. We use the generated random density matrix elements as inputs for the event rate calculations performed with the \
\texttt{dmscatter} code~\cite{Gorton:2022eed}. 
{\color{black}  We emphasize that 
this is a preliminary foray into 
uncertainty quantification (UQ) of the nuclear input into DM direct detection. Our results, given below, provide motivation for future investigations.}

\section{Results for Ideal Event Rate Spectra}
\label{sec:results}

This section presents results for the ideal DM-nucleus differential event rates for xenon isotopes and their uncertainties calculated using the methods described in the previous section.

\begin{figure*}[t]
    \centering
    \includegraphics[width=\linewidth]{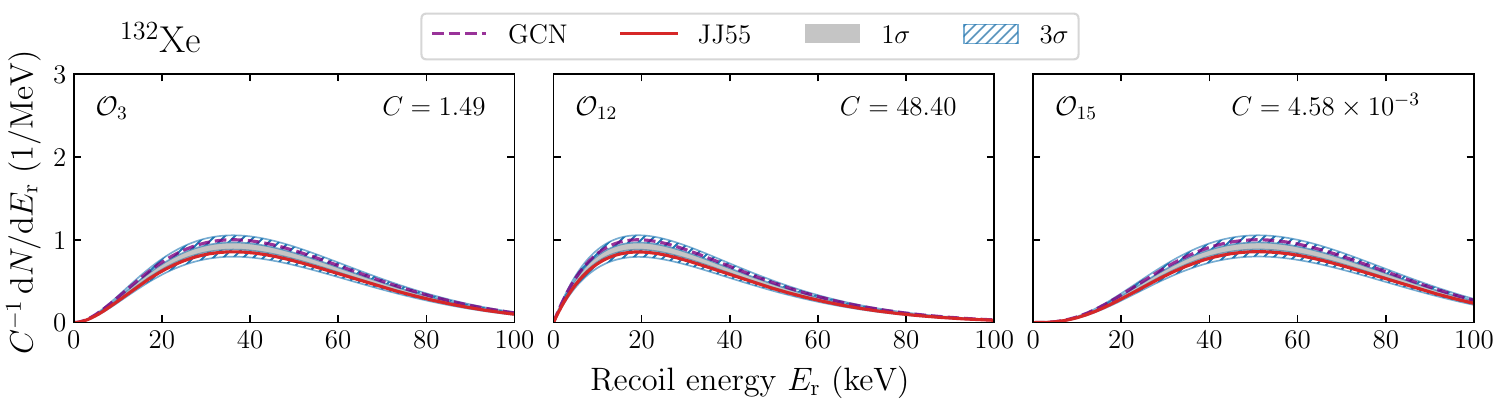}\\
    \vspace{2em}
    \includegraphics[width=\linewidth]{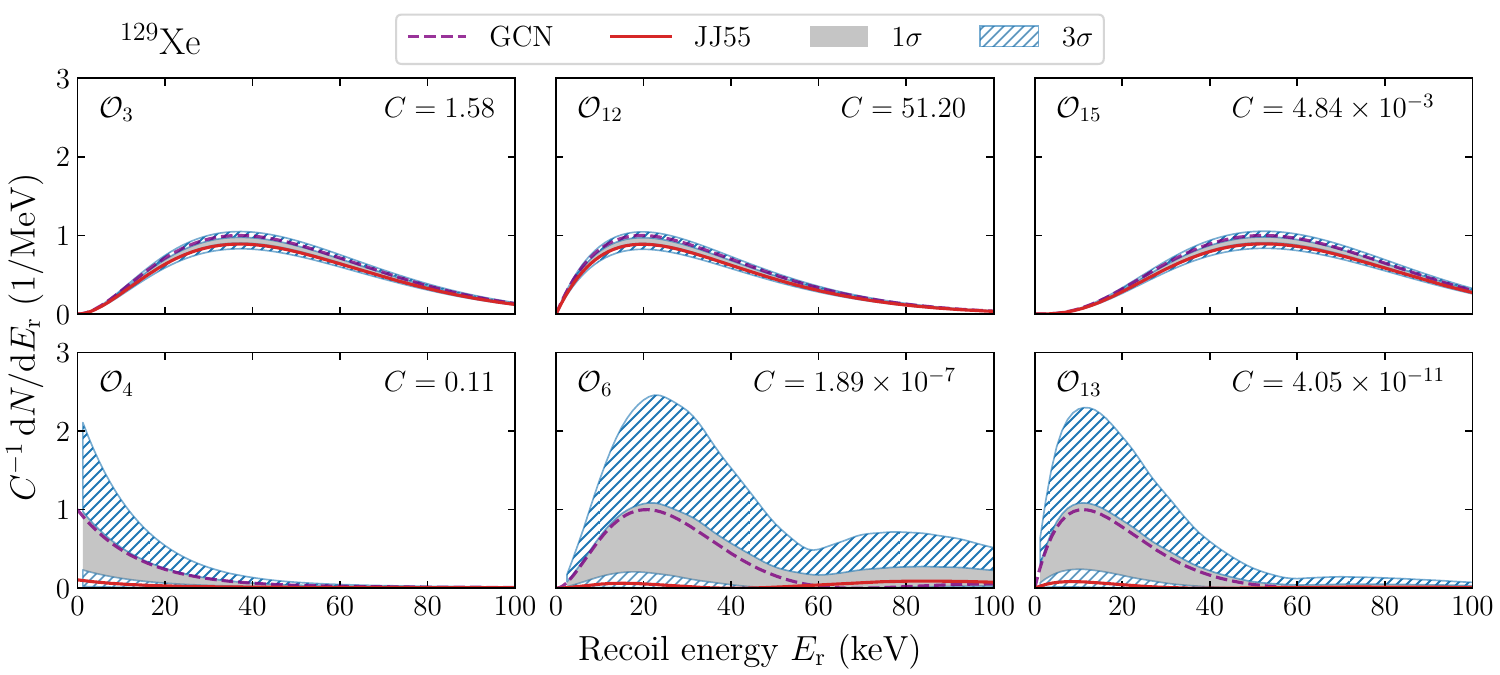}
    \caption{Spread in the differential event rates $\mathrm{d}N/\mathrm{d}E_\text{r}$ for WIMPs interacting through DM nonrelativistic EFT operators $\mathcal{O}_3$, $\mathcal{O}_{12}$, and $\mathcal{O}_{15}$ with $^{132}\text{Xe}$ (top three plots), and $\mathcal{O}_3$, $\mathcal{O}_{12}$, $\mathcal{O}_{15}$, $\mathcal{O}_4$, $\mathcal{O}_6$, and $\mathcal{O}_{13}$ with $^{129}$Xe (bottom six plots), assuming an effective exposure of 1~ton~yr, ideal efficiency $\varepsilon(E_\text{r})=1$, WIMP mass $m_\chi = 150$~GeV, and DM EFT coupling to protons $c_i^{\mathrm{p}}m_v^2 = 0.25$, where $m_v~=~246.2$~GeV is the weak interaction mass scale. The plots demonstrate differences in the magnitude and shape of the scattering rates of WIMPs with Xe nuclei. Each panel is arbitrarily normalized such that the maximum of the GCN event rate (orange line) is 1/MeV, and the corresponding normalization factor $C$ is provided in the top right of each plot. The actual ideal event rate is $C$ multiplied by the event rate shown.}
    \label{fig:event-rates}
\end{figure*}

Figure~\ref{fig:event-rates} shows the 1$\sigma$ and 3$\sigma$ uncertainties in event rate spectra for WIMP scattering off $^{132}\text{Xe}$ and $^{129}\text{Xe}$ in the proton channel for a selection of EFT operators, assuming a WIMP mass of $m_\chi = 150$~GeV.
Each of the event rate spectra subplots was generated with $N=1000$ input files containing randomized nuclear density matrix elements produced from the Monte Carlo method described in the previous section.
Several of the EFT operators yield significantly different event rates for the two shell models used. These large spreads translate into high uncertainties in the event rates. 

Figure~\ref{fig:event-rates} emphasizes how the event rate spectra can differ significantly in shape and magnitude for various DM EFT operators. For example, the lower left and lower middle panels show the event rate spectra for operators $\mathcal{O}_4$ - Eq.~\eqref{eq:O4} and $\mathcal{O}_6$ - Eq.~\eqref{eq:O6} in $^{129}$Xe. The former operator is the simple spin-dependent coupling where low energy recoils are favored (as in the case of coherent scattering), whereas the latter operator results in a peak in the spectrum due to an additional power of $(\vec{q} / m_N)^2$ present in the cross section.

As mentioned, for many of the EFT operators, (see, e.g., the aforementioned $\mathcal{O}_4$, $\mathcal{O}_{6}$ in Fig.~\ref{fig:event-rates},) the two configuration-interaction Hamiltonians GCN and JJ55 predict vastly different event rates, particularly in the low recoil energy regime; {\color{black} we hypothesize such differences are due to  cancellations in Eq.~(\ref{eq:density-matrix}).} As such, the exclusion limits on the WIMP-nucleon coupling for these operators or the potential discovery fits will be sensitive to the choice of a given shell model interaction.

The mass of the WIMP affects the ‘length’ of the event rate spectra, i.e., where the event rate spectra tapers
off to zero. As the WIMP mass increases, the WIMP carries more momentum and is thus able to cause collisions at higher nuclear recoil energies. This has implications for the discovery potential for DM detectors, as a ‘longer’ event rate spectrum would allow for more potential events if the detector is capable of registering events at higher nuclear recoil energies. On the other hand, a low WIMP mass would correlate to a ‘short’ event rate spectrum; if a detector is unable to probe scatterings at low recoil energies, it is possible that the majority of the WIMP-nuclei scatterings will go unseen, as the detector will be searching a recoil energy range that is mostly unpopulated.

\section{Sensitivity analysis for XENON1T}
\label{sec:detector-limits}

So far there has been no conclusive detection of particle DM by a direct DM detection program. Therefore, only upper limits on WIMP coupling strengths to Standard Model particles have been set. 

In this section, we describe how the process detailed in Sec.~\ref{sec:Tools-and-Methods} can be used to calculate the uncertainty coming from nuclear shell models on the upper limits of WIMP couplings to nucleons. For the remainder of this work, we use the XENON1T experiment~\cite{XENON:2018voc} as an example, although the uncertainty quantification procedure we will describe henceforth will work for any detector/element, provided that more than one nuclear density matrix data set is available.

XENON1T~\cite{XENON:2015gkh, XENON:2018voc} was a 3.2 ton dual-phase xenon time projection chamber DM direct search experiment located at the Laboratori Nazionali del Gran Sasso in Italy. The separate detection of the prompt scintillation in the liquid phase of the detector (S1 signal) and delayed scintillation in the gaseous phase of the detector (S2 signal) enabled a powerful discrimination between the electronic and nuclear recoils~\cite{Aprile:2006kx, XENON:2022avm}. As such, to good approximation the only backgrounds for the DM-nuclei interactions are those producing nuclear recoils only, which are close to zero in the 1~ton~yr effective exposure of XENON1T~\cite{XENON:2018voc, XENON:2022avm}.

To get the realistic upper limits on the DM EFT coupling coefficients we use the 1~ton~yr effective exposure and detection efficiency $\varepsilon(E_\text{r})$ given by the curve in Ref.~\cite{XENON:2022avm} used for the spin-independent analysis. We restrict our analysis to the region of interest in the recoil energy of $E_\text{r} = [4.9, 54.4]$~keV.
We calculate the total number of events in 1~ton~yr exposure of XENON1T detector integrating the differential event rate Eq.~\eqref{eq:diffeventrate} over the recoil energy window of the detector with the specified efficiency.

With the number of events per effective exposure, one can calculate the 90\% confidence level (90\% C.L.) upper limits on the EFT couplings of the DM particle. The 90\% C.L. represents the smallest coupling, given a set of detector characteristics, at which a statistically significant number of WIMP-nuclei scatterings would be detected. Carrying this calculation out for varying WIMP mass $m_\chi$ traces a curve in the $c_i^x$~--~$m_\chi$ plane. The region of phase space above this curve represents all $(c_i^x, m_\chi)$ combinations that through experimental non-detection have been excluded as realistic scenarios, while the region below represents combinations that have yet to be probed.

To calculate these exclusion curves (upper limits), we assume that XENON1T is a zero background experiment, based on the successful discrimination of background signals in the S1 vs. S2 phase space plane~\cite{XENON:2018voc, XENON:2022avm}. To model the detector composition, we use the abundance-weighted average of differential event rates over the six most abundant xenon isotopes ($^{132}\text{Xe}$, $^{129}\text{Xe}$, $^{131}\text{Xe}$, $^{134}\text{Xe}$, $^{136}\text{Xe}$, and $^{130}\text{Xe}$, in that order). The relative abundances of these isotopes by mass are respectively 26.9\%, 26.4\%, 21.2\%, 10.4\%, 8.86\%, and 4.07\%; when summed, these six isotopes constitute slightly more than $97\%$ of the mass of the XENON1T detector. There is a different amount of uncertainty that each isotope contributes, which is discussed further in Sec.~\ref{sec:Discussion}.

\begin{figure*}[t]
    \centering
    \includegraphics[width=\linewidth]{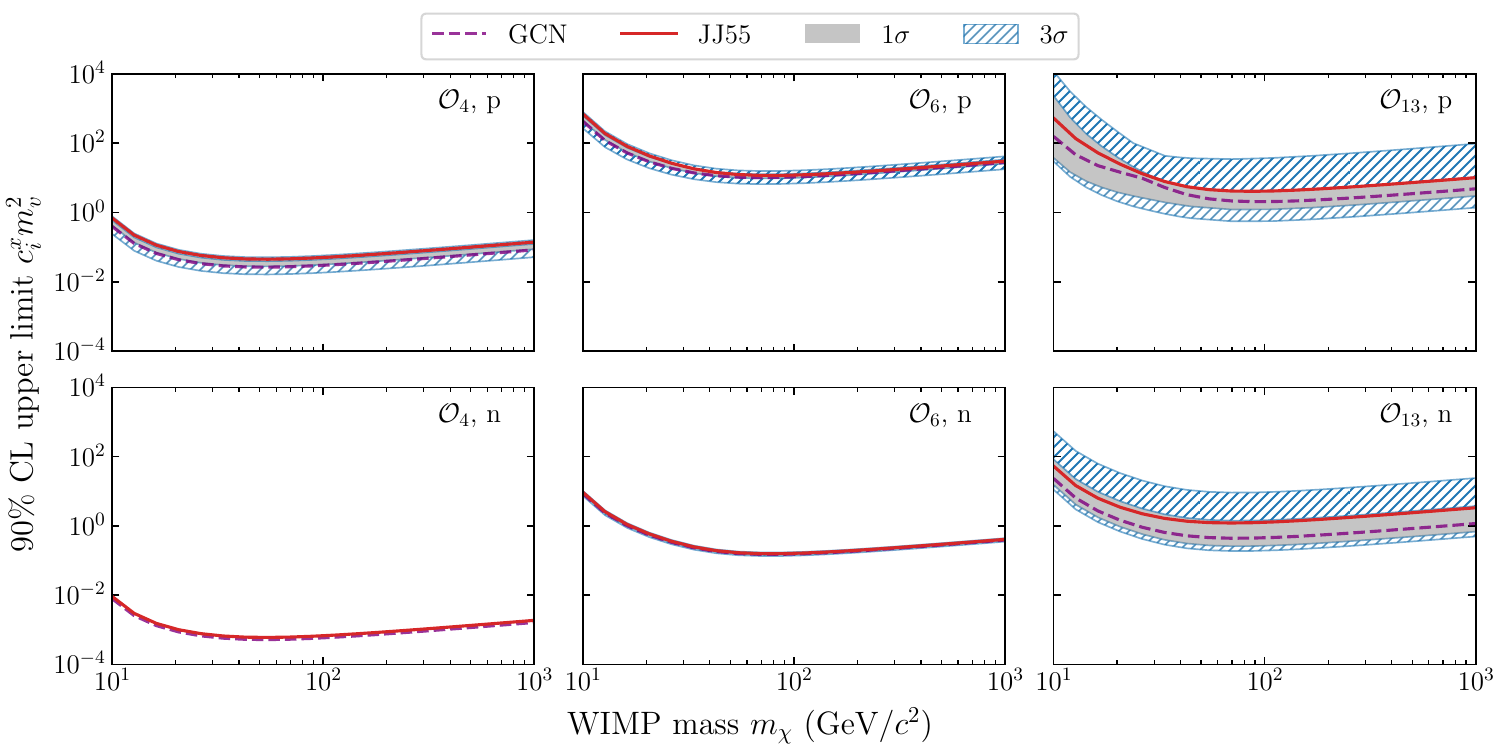}
    \caption{The 90\% C.L. upper limits on the nonrelativistic EFT coupling coefficients (as detailed in Sec.~\ref{sec:detector-limits}) for select nucleon and operator combinations along with $1\sigma$ and $3\sigma$ uncertainty bands from nuclear uncertainties calculated using XENON1T experiment as a model for a general DM direct detection experiment. The top three plots are exclusion plots for the EFT couplings to protons, while the bottom three plots are exclusion plots for the EFT couplings to neutrons. We use a common vertical scale for all of the plots to show which operators are more constrained than others. Note that the vertical scale obscures the non-zero sensitivity of $\mathcal{O}_{4}$ Eq.~\eqref{eq:O4} to the variations in nuclear density matrices (bottom left plot).}
    \label{fig:sampleExclusionPlots}
\end{figure*}

The event rate $dN/dt$ is linearly dependent on the dark matter response functions $R_X^{x,x'}$ through the cross section. Further, if we assume isospin is conserved $x=x'$ (i.e., we only look at p$\rightarrow$p and n$\rightarrow$n interactions as is the case in the ground state-to-ground state interaction), the $R_X^{x,x'}$ depend quadratically on the EFT coefficients. If only a single EFT coefficient $c_i^x$ is nonzero then the total event rate for that coupling is a function of $(c_i^x)^2$. Thus, to find the minimal coupling needed to produce a zero background event rate $dN_{\text{min}}/dt$ required for a statistically significant WIMP-nucleus scattering detection, we only need to run the above event rate procedure once for a test coupling $c_{i,0}^x$. Assuming the number of WIMP-nucleon scattering events follows the Poisson distribution, in order to achieve a 90\% confidence of detection we need an event rate of at least $dN_{\text{min}}/dt = 2.3$~events~ton$^{-1}$~yr$^{-1}$ in the zero background case. The chosen test coupling $c_{i,0}^x$ yields an event rate of $dN_0/dt$, which we then use to scale the EFT coefficient to find the minimal required coupling for detection, $c_{i,\text{min}}^x$:
\begin{equation}
\label{eq:ci-scaling}
    c_{i,\text{min}}^x(m_\chi) = \sqrt{\frac{dN_{\text{min}}/dt}{dN_0/dt}} ~c_{i,0}^x \ .
\end{equation}
By calculating the minimal couplings for WIMP masses in the range $10 ~\text{GeV} \leq m_\chi \leq 1000 ~\text{GeV}$, we create exclusion curves for each EFT coefficient as a function of $m_\chi$ for repeated calculations with randomized nuclear density matrices. We ran $N_\mathrm{trial} = 1000$ MC trials for most of EFT coefficients for each $m_\chi$\footnote{For $\mathcal{O}_{13}$, the distribution of exclusion curves was highly non-Gaussian, and as such we increased $N_\mathrm{trials}$ to $10^4$ in order to obtain a more accurate probability distribution.}. Thus, we can quantify the uncertainty on the coupling limits that derive from the uncertainties in the nuclear models from the spread in the $N_\mathrm{trial}$ exclusion curves.

The distributions of the exclusion curve values at each $m_\chi$ are in general not Gaussian despite the nuclear matrix elements in the MC simulation being drawn from normal distributions. To calculate the $1\sigma$ bands, we use the Feldman-Cousins procedure described in Ref.~\cite{Feldman:1997qc}. We find the interval in coefficient values $[a,b]$ such that the integral of the histogram $H(c_i; m_\chi)$ of coefficient values $c_i$ at a given $m_\chi$ satisfies
\begin{equation}
\label{eq:probability-distribiution-integral}
    \frac{\int_a^b dc_i\, H(c_i; m_\chi)}{\int_{-\infty}^{\infty} dc_i\, H(c_i; m_\chi)} = 0.68 \ ,
\end{equation}
and $H(a; m_\chi) = H(b; m_\chi)$. Figure~\ref{fig:example_1sigma_distribution} in App.~\ref{app:coupling-distribution} shows an example of such a $1\sigma$ band for $c_{13}^p$ at $m_\chi = 12$~GeV.

\begin{figure*}
    \includegraphics[width=0.99\linewidth]{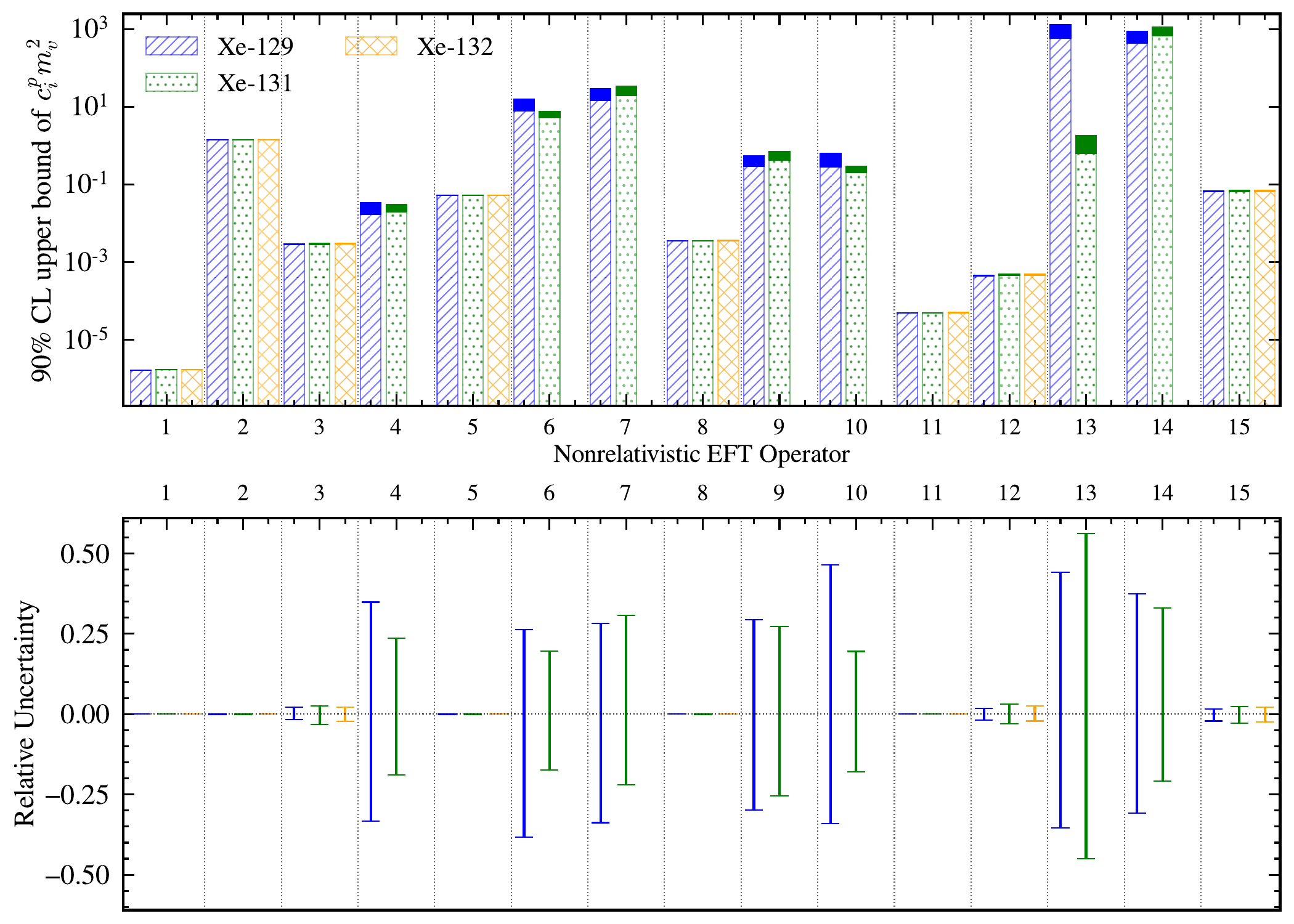}
    \caption{The 90\% C.L. upper limits on the proton-WIMP nonrelativistic EFT coupling coefficients (top) and their relative uncertainties (bottom) for $^{129}$Xe, $^{131}$Xe, and $^{132}$Xe, assuming a WIMP mass of $m_\chi = 70$~GeV. Not shown are the three other even-even isotopes as they have nearly identical sensitivities to $^{132}$Xe for all EFT channels. Note that xenon isotopes with even mass number are not sensitive to operators $\mathcal{O}_4$, $\mathcal{O}_6$, $\mathcal{O}_7$, $\mathcal{O}_9$, $\mathcal{O}_{10}$, $\mathcal{O}_{13}$, and $\mathcal{O}_{14}$.}
    \label{fig:isotope_uncert}
\end{figure*}

\begin{figure*}
    \includegraphics[width=0.99\linewidth]{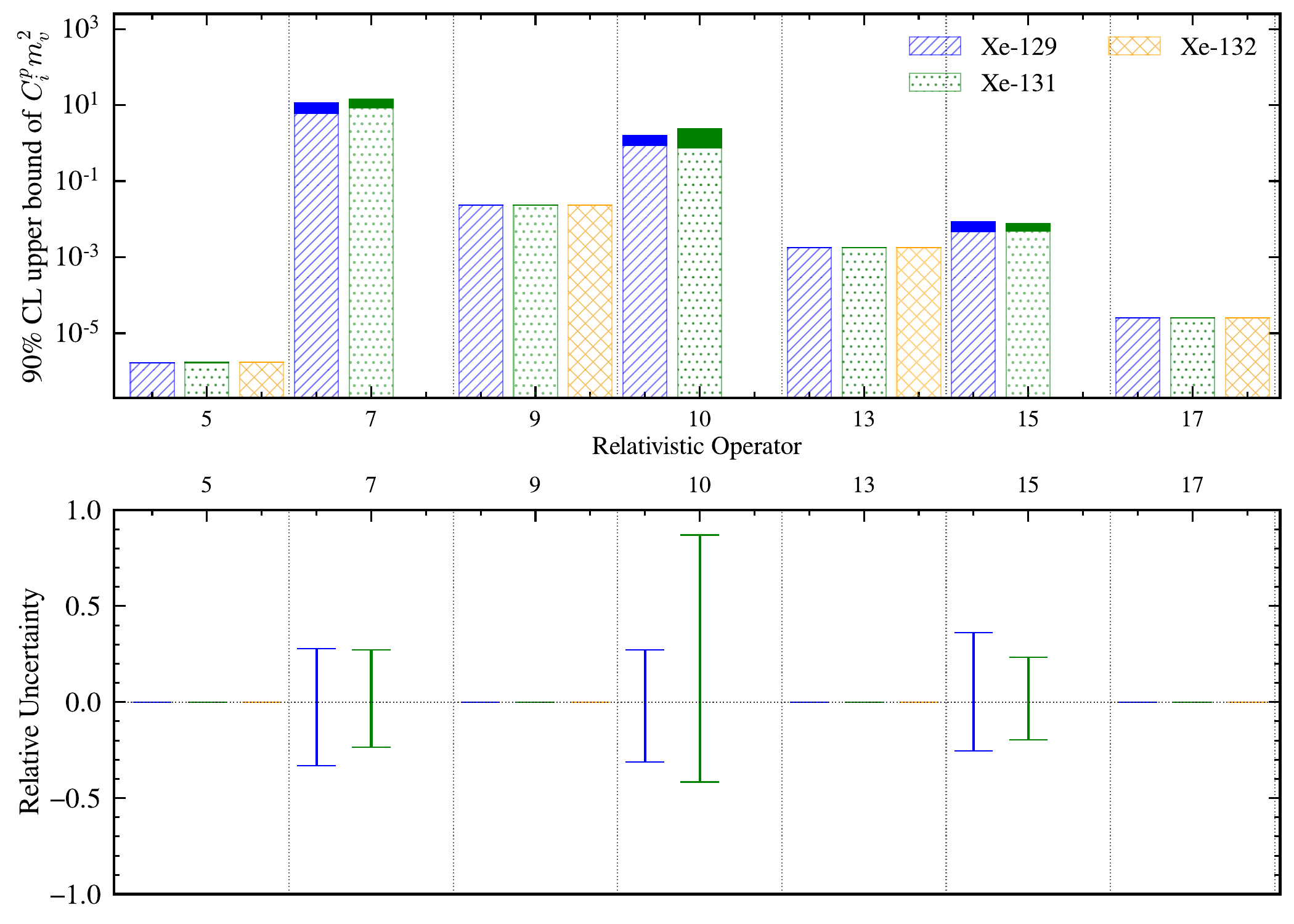}
    \caption{The 90\% C.L. upper limits on the proton-WIMP relativistic couplings (top) and their relative uncertainties (bottom) for $^{129}$Xe, $^{131}$Xe, and $^{132}$Xe, assuming a WIMP mass of $m_{\chi} = 69.5$~GeV. Definitions for the relativistic couplings in terms of the nonrelativistic EFT operators can be found in Eq.~(\ref{eq:rel_operators}). Note that even-even xenon isotopes are not sensitive to operators $\mathcal{L}_7^{\text{int}}$, $\mathcal{L}_{10}^{\text{int}}$, and $\mathcal{L}_{15}^{\text{int}}$.}
    \label{fig:isotope_rel_uncert}
\end{figure*}


Figure~\ref{fig:sampleExclusionPlots} shows the calculated examples of the upper limits on the DM-nuclei coupling coefficients for protons (upper panels) and neutrons (lower panels) as a function of WIMP mass and their uncertainty for selected DM EFT operators calculated for the XENON1T detector. The remaining combinations that exhibit sensitivity to the shell-model uncertainties are presented in App.~\ref{app:exclusion-curves} for completeness.

The WIMP-nuclei upper coupling limit is noticeably more sensitive to the density matrices when coupling through operator $\mathcal{O}_{13}$ in Eq.~\eqref{eq:O13} than other EFT DM operators for both protons and neutrons, corresponding to potentially larger uncertainty. Notably, for most of the neutron EFT operators, the exclusion curves generated using our MC process (see Sec.~\ref{sec:Tools-and-Methods}) are normally distributed. However, the distribution for $\mathcal{O}_{13}$ is not normally distributed and is skewed toward lower couplings. Operators $\mathcal{O}_9$ and $\mathcal{O}_{14}$ for neutrons and $\mathcal{O}_4$, $\mathcal{O}_6$, $\mathcal{O}_9$, $\mathcal{O}_{10}$, and $\mathcal{O}_{14}$ for protons also exhibit relatively greater sensitivity for both proton and neutron couplings.

\section{Discussion}
\label{sec:Discussion}

Our results demonstrate that the need to implement nuclear models that involve approximations can lead to substantial uncertainties in the interpretation of the results of dark matter detection experiments.

Figure~\ref{fig:isotope_uncert} shows the upper limits on the nonrelativistic proton-WIMP coupling coefficient values and their relative uncertainties for $^{129}$Xe, $^{131}$Xe and $^{132}$Xe assuming $m_\chi = 70$~GeV, and the total fiducial mass of XENON1T is comprised of each isotope, respectively. We show here the results only for the most abundant even-even isotope of xenon because all four even-even isotopes had nearly identical sensitivities and uncertainties for each EFT coefficient and did not probe operators $\mathcal{O}_4$, $\mathcal{O}_6$, $\mathcal{O}_7$, $\mathcal{O}_9$, $\mathcal{O}_{10}$, $\mathcal{O}_{13}$, or $\mathcal{O}_{14}$. {\color{black}Standard selection rules for parity and angular momentum cause the matrix elements of these operators to vanish for $J_i=0$ ground states.} 

The two even-odd isotopes, $^{129}$Xe and $^{131}$Xe, have $J_i >0$ ground states and thus probed all operators while producing more uncertain upper limits in most cases for the operators present for even-even isotopes.  

Although the  dark matter-nucleon couplings were derived from Galilean-invariant effective field theory~\cite{Anand:2013yka}, one can also start from a relativistic, Lorentz-invariant Lagrangian, including those derived from chiral effective field theory, and carry out a nonrelativistic reduction~\cite{Anand:2013yka,PhysRevD.94.063505,hoferichter2015chiral,bishara2017quarks,xia2019pandax,hoferichter2019darkmatternucleus,PhysRevD.99.055031,PhysRevC.106.044003} of the WIMP-nucleon interaction terms. This leads to specific linear combinations of the nonrelativistic operators, with possible additional dependencies upon the momentum transfer $q$ and WIMP mass $m_{\chi}$. We considered the seven cases given in~Ref.~\cite{xia2019pandax}:
\begin{widetext}
\begin{subequations}
\label{eq:rel_operators}
\begin{eqnarray}
{\cal L}^\mathrm{int}_5 
\equiv  \bar{\chi} \gamma^\mu \chi \, \, 
\bar{N} \gamma_\mu N & \rightarrow  &
{\cal O}_1; \\
{\cal L}^\mathrm{int}_7 
 \equiv  \bar{\chi} \gamma^\mu \chi \, \, 
\bar{N} \gamma_\mu \gamma^5 N  &\rightarrow &-2
{\cal O}_7 + 2 \frac{m_N}{m_\chi} {\cal O}_9; \\
{\cal L}^\mathrm{int}_{9} 
 \equiv  \bar{\chi} i \sigma^{\mu \nu}
\frac{ q_\nu}{m_M}\chi \, \, 
\bar{N} \gamma_\mu N  &\rightarrow &- \frac{\vec{q}^{\:2}}{2 m_\chi m_M } {\cal O}_1
+ 2 \frac{ m_N}{m_M}{\cal O}_5 
- 2 \frac{ m_N}{m_M} \left ( \frac{\vec{q}^{\:2}}{m_N^2 } {\cal O}_4
-{\cal O}_6\right );
 \\
 {\cal L}^\mathrm{int}_{10} 
 \equiv  \bar{\chi} i \sigma^{\mu \nu}
\frac{ q_\nu}{m_M}\chi \, \, 
\bar{N} i \sigma_{\mu \alpha}
\frac{ q^\alpha}{m_M} N  & \rightarrow &
4 \left (\frac{\vec{q}^{\:2}}{m^2_M } {\cal O}_4 -  
\frac{m_N^2}{m^2_M } {\cal O}_6 \right );
\\
{\cal L}^\mathrm{int}_{13}
 \equiv  \bar{\chi} \gamma^\mu \gamma^5 \chi \, \, 
\bar{N} \gamma_\mu N  &\rightarrow & 2
{\cal O}_8 + 2 {\cal O}_9; \\
{\cal L}^\mathrm{int}_{15} 
 \equiv  \bar{\chi} \gamma^\mu \gamma^5 \chi \, \, 
\bar{N} \gamma_\mu \gamma^5 N  &\rightarrow &-4 {\cal O}_4;\\
{\cal L}^\mathrm{int}_{17} 
 \equiv  i \bar{\chi} i \sigma^{\mu \nu}
\frac{ q_\nu}{m_M} \gamma^5 \chi \, \, 
\bar{N} \gamma_\mu N & \rightarrow &
 2 \frac{ m_N}{m_M}{\cal O}_{11};
\end{eqnarray}
\end{subequations}
\end{widetext}
where $\chi$ is the WIMP four-spinor (and thus one must assume the WIMP is spin $1/2$) and 
$N$ the nucleon spinor; $m_M$ defines 
the scale of the effective theory, often chosen to be either the weak interaction 
scale, $m_v \approx 246$ GeV~~\cite{Anand:2013yka}, or the nucleon mass~\cite{xia2019pandax}.

The linear combination of the Galilean EFT operators allows for potential interference between operators when calculating event rates. Coupled with the momentum transfer and $m_{\chi}$ dependence, uncertainties in minimal relativistic couplings $C_{i}^{x}$ for the relativistic interaction terms could have a non-trivial relation to the minimal coupling uncertainties of the Galilean EFT operators. Thus, we repeated the analysis of the uncertainty in minimal coupling coefficients (see Secs.~\ref{sec:Tools-and-Methods} - \ref{sec:detector-limits} and Fig. \ref{fig:isotope_uncert}) for the seven relativistic operators $\mathcal{L}^{\text{int}}_i$ above. To do this, we modified the {\tt dmscatter} code~\cite{Gorton:2022eed} to specify these relativistic couplings. 

Figure~\ref{fig:isotope_rel_uncert} shows the 90\% confidence level minimal coupling coefficients and their relative uncertainties for the relativistic interaction couplings. As expected, terms that are combinations of Galilean EFT operators with negligible uncertainties (i.e., $\mathcal{L}_{5}^{\text{int}}$, $\mathcal{L}_{9}^{\text{int}}$, $\mathcal{L}_{13}^{\text{int}}$, and $\mathcal{L}_{17}^{\text{int}}$) have negligible uncertainties derived from the nuclear structure uncertainties.
{\color{black} In general, the uncertainties do not reflect significant interference between the nonrelativistic operators; the one exception is} interaction term 10, which has a relative uncertainty in $^{131}$Xe that  exceeds the uncertainty, {\color{black} by a factor of roughly 3}, in either of the Galilean EFT operator minimal couplings that it depends on, namely $\mathcal{O}_4$ and $\mathcal{O}_6$.
{\color{black} Therefore, our conclusions are not very sensitive to the choice of relativistic or nonrelativistic formalism.}

{\color{black} Figures \ref{fig:isotope_uncert} and \ref{fig:isotope_rel_uncert} suggest that terms containing nuclear spin ($\vec{S}_N$ in Eqns.~\eqref{eq:operators} and nucleon axial vector or tensor terms in Eqns.~\eqref{eq:rel_operators}) have the largest uncertainties stemming from nuclear model uncertainties. In Appendix \ref{app:nuclear-tests} we compare energies and electromagnetic observables 
from our two models to experimental results for pertinent xenon isotopes.  
Among the observables are magnetic dipole moments, which include an effective coupling to nucleon spins.
We get reasonable agreement, with neither model systematically better than the other; is most cases the calculated electromagnetic observables are within experimental error.  
This bolsters confidence in our ability to model the nuclear response to WIMP scattering, including couplings dominated by nucleon spin, as well as to estimate theoretical uncertainties. 

}

Uncertainties from nuclear modeling are obviously not the only source of error in dark matter detection experiments. 
For example, 
recoil rates can be heavily suppressed by isospin-violating DM interactions in DM direct detection experiments~\cite{alanne2022z,cheek2023isospinviolating}.
Similarly, Ref.~\cite{Fowlie:2018svr} showed that non-parametric uncertainties in the dark matter halo velocity distribution could weaken exclusion limits in XENON1T by as much as two orders of magnitude, although this assumes $m_\chi \leq 60$~GeV and anisotropy in the dark matter halo. Above $m_\chi = 60$~GeV, uncertainties have a mild effect on the limits. 
Therefore, there is a reasonable parameter space in which the nuclear uncertainties studied in this paper could dominate the overall uncertainty of the experiments.

\subsection{Strategies for improving the calculations}
\label{sec:quantum-computing-discussion}

The methods used to obtain the results presented here could be improved both by the use of more sophisticated but computationally intensive methods to estimate the uncertainties and by developing methods that enable the uncertainties themselves to be reduced.

Regarding our estimates of the uncertainties, it is worth emphasizing that we have modeled the uncertainty in the nuclear shell-model by creating variations in the reduced one-body density matrix elements, as shown in the center column of Fig.~\ref{fig:flowchart}. A more precise uncertainty analysis would begin by estimating the uncertainties in the nuclear Hamiltonian matrix elements~\cite{fox_usdb} (top-left box of Fig.~\ref{fig:flowchart}), and propagating this uncertainty~\cite{fox2022uncertainty} through the same flowchart from there, ultimately to the minimal EFT couplings again. The downside to this alternate method is that the configuration-interaction calculations used to reduce the many-body system to the more tractable one-body density matrices (see left column of Fig.~\ref{fig:flowchart}) are computationally intensive as well as time-consuming. 
{\color{black} One potential alternative is the use of emulators~\cite{konig_ec}, but the application to such shell-model calculations has yet to be performed.}

{\color{black} Much of the challenge of modeling the nuclear response to DM is the daunting dimensions, to $10^{10}$ or beyond, of even a truncated Hilbert space.}
Improving the uncertainty estimates and reducing the uncertainties of the calculations themselves are both areas in which quantum computing has the potential to excel.
We note, however, the large number of nucleons in the relevant nuclei and the computational depth of the necessary computations are such that achieving higher accuracies than classical methods would require substantial improvements in quantum computing hardware and error correction.
In any case, the preliminary analysis that we have presented here can highlight which isotope and EFT operator combinations exhibit nontrivial uncertainty and therefore are attractive targets for further investigation.

\section{Conclusions}
\label{sec:Conclusions}

Understanding the nature of dark matter is critical to progress in physics, as it could shed light on some of the most fundamental questions in cosmology and particle physics. While there has been significant progress in direct dark matter detection experiments, so far only gravitational interactions of dark matter have been observed. Direct detection experiments aim to observe the interactions of particle dark matter with ordinary matter, which could provide valuable information about its properties, such as mass, type of the interaction, and interaction strength. This information can help in determining the underlying theory of dark matter, which could have significant implications for our understanding of the universe at large.

We have presented a preliminary uncertainty analysis of the WIMP-nucleon coupling exclusion limits stemming from uncertainties in nuclear shell model calculations. We have applied this `first-order' analysis to study the sensitivity of all operators in the Galilean effective field theory, and identified the EFT operators (couplings) which are particularly affected by current limitations of nuclear shell model calculations. The analysis presented here could also be easily applied to direct detection experiments using different elements in the detector, such as argon (DarkSide-50~\cite{DarkSide-50:2020swd}) or tellurium (CUORE~\cite{ADAMS2022103902}).
We also calculated the change in the uncertainties on the coupling limits for the operators obtained by performing nonrelativistic reductions of the relativistic interactions.

Our analysis reveals that the degree of uncertainty in the upper limits of the coupling coefficients varies considerably across the different Galilean dark matter Effective Field Theory operators. Specifically, we observe that the uncertainties for the upper limits on couplings to certain operators can be substantial, with the $\pm1\sigma$ range spanning over an order of magnitude in the limit of low dark matter masses below approximately 20~GeV for the operator $\mathcal{O}_{13}$. These findings underscore the importance of accounting for nuclear uncertainties when determining the upper limits on dark matter-nucleus coupling coefficients in direct dark matter experiments. Moreover, the advancements in the precision of nuclear structure calculations have the potential to greatly enhance the accuracy of results obtained from dark matter detection experiments.

\begin{figure*}[t]
    \centering
    \includegraphics[width=\linewidth]{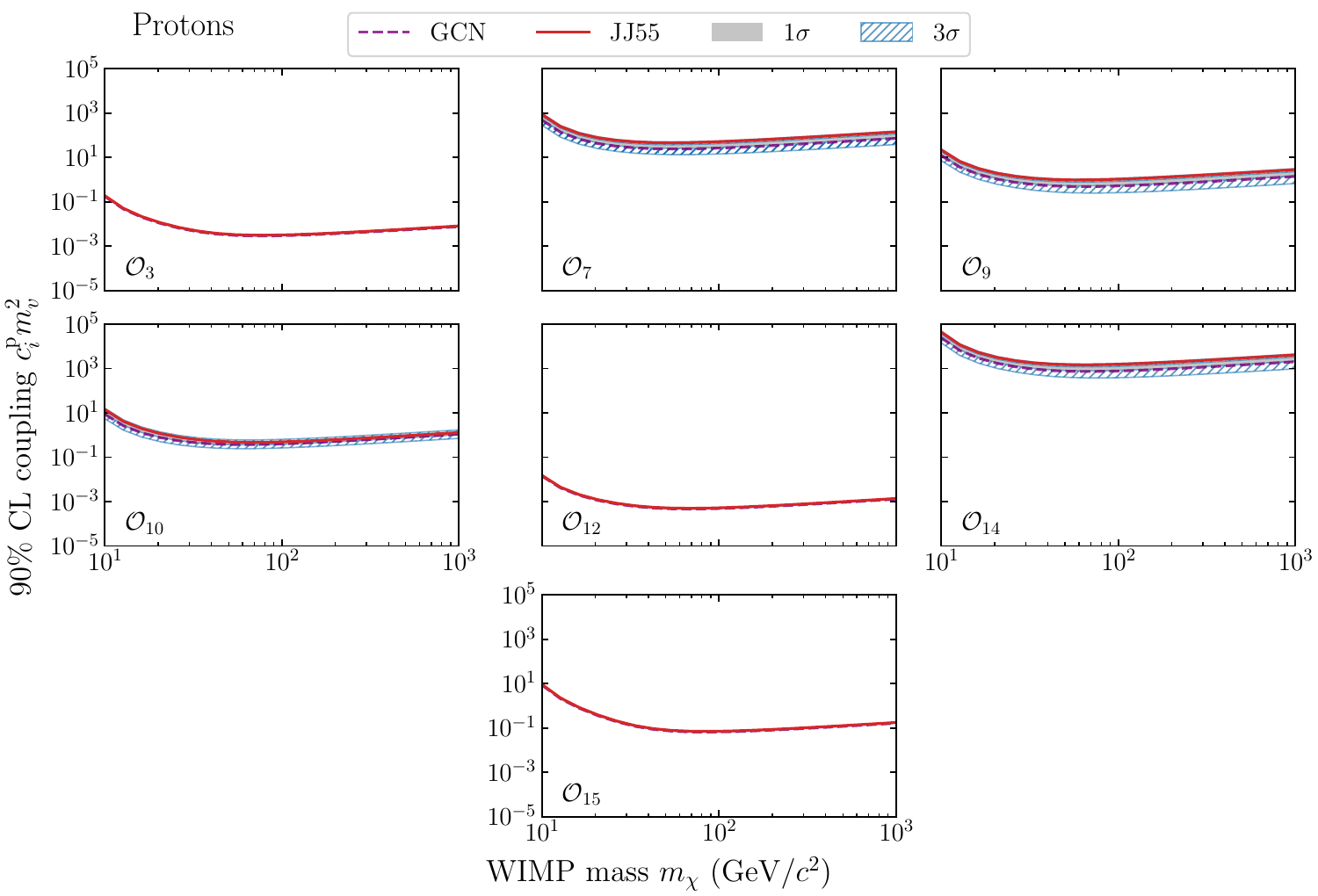}\\
    \vspace{2em}
    \includegraphics[width=\linewidth]{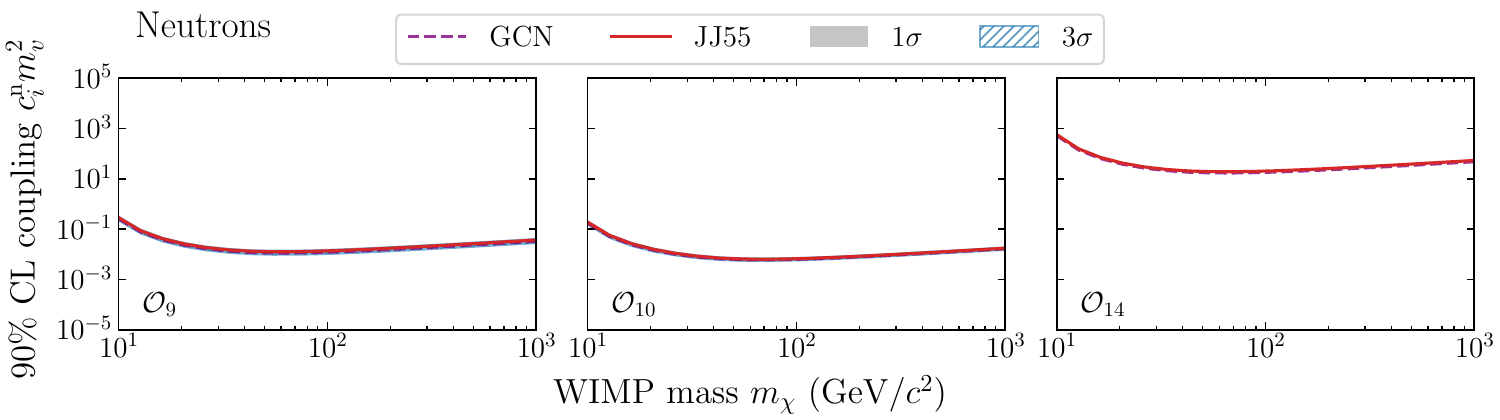}
    \caption{The 90\% C.L. upper limits on the nonrelativistic EFT coupling coefficients (as detailed in Sec.~\ref{sec:detector-limits}) for select proton and operator combinations along with $1\sigma$ and $3\sigma$ uncertainty bands from nuclear uncertainties calculated using XENON1T experiment as a model for a general DM direct detection experiment.
    Note that the vertical scale obscures the non-zero sensitivity of $\mathcal{O}_3$ Eq.~\eqref{eq:O3}, $\mathcal{O}_{12}$ Eq.~\eqref{eq:O12} and $\mathcal{O}_{15}$ Eq.~\eqref{eq:O3} to the variations in nuclear density matrices.}
    \label{fig:exclusion_protons}
\end{figure*}

\begin{acknowledgments}

We are grateful for helpful discussions with Oliver Gorton, Wick Haxton, Evan Rule, and Pooja Siwach.
This work was supported in part by the U.S.~Department of Energy, Office of Science, Office of High Energy Physics, under Award No.~DE-SC0019465, in part by National Science Foundation Grants PHY-2020275 and PHY-2108339, and in part by an unrestricted gift to UNSW Sydney by Google AI.
A.M.S. and A.B.B thank the Kavli Institute for Theoretical Physics (KITP) for hospitality during this work. Work at KITP was supported in part by the National Science Foundation under Grant No. NSF PHY-1748958. A.M.S.\ acknowledges the Institute for Nuclear Theory (INT) at the University of Washington for partial support and for its kind hospitality and stimulating research environment. Work at INT was supported in part by the INT's U.S. Department of Energy grant No. DE-FG02-00ER41132.

\end{acknowledgments}

\appendix

\begin{figure}
    \centering
    \includegraphics[width=\columnwidth]{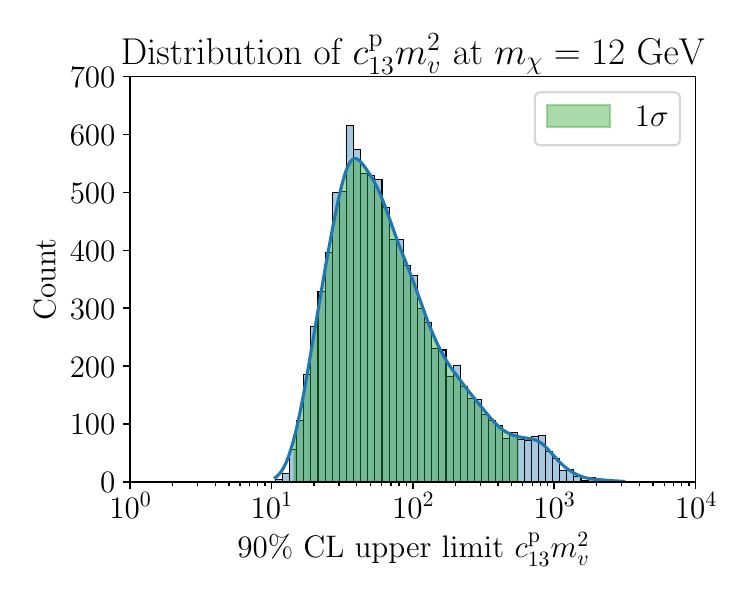}
    \caption{Example distribution of the 90\% C.L. upper limit on $c_{13}^\mathrm{p}$ for $N_\text{trials}=10^4$ MC density matrices at a WIMP mass of $m_\chi=12$ GeV.
    The shaded green region represents the $1\sigma$ band and contains 68\% of the distribution as found from the Feldman-Cousins procedure as explained in Sec.~\ref{sec:detector-limits}. The $3\sigma$ band, containing 99.7\% of the distribution, is approximately represented by the support of the histogram. The obtained distribution is highly non-symmetric.}
    \label{fig:example_1sigma_distribution}
\end{figure}
\section{The upper limits on the coupling coefficients for all sensitive operators}
\label{app:exclusion-curves}

In this appendix we include the calculated exclusion plots for XENON1T, same as shown in Fig.~\ref{fig:sampleExclusionPlots}, for the sensitive operator and nucleon combinations not presented in the main text. In total, there are 30 possible nonrelativistic channels (15 EFT operators and two nucleons) for interaction; however, we do not include the plots for operator and nucleon combinations with next-to-no uncertainty in this appendix. Some channels exhibit negligible uncertainty in the event rate due to the nuclear density matrices. Once we calculate the exclusion curves for these channels, with the procedure described in Sec.~\ref{sec:detector-limits}, the already small uncertainty becomes even more insignificant. In contrast, $\mathcal{O}_{13}$ exhibits significant uncertainty (variation) in the event rate, both in amplitude and features (shape, local minima and maxima, etc.).  With different shape and amplitude in the event rate, the exclusion curves calculated for $\mathcal{O}_{13}$ using the two odd-even isotopes of xenon have equal, if not greater, uncertainty than the event rate curves themselves.

Figure~\ref{fig:exclusion_protons} shows the calculated examples of the upper limits on the DM-nuclei couplings for protons (upper panels) and neutrons (lower panels) as a function of WIMP mass and their uncertainty for and EFT operators calculated for the XENON1T detector as described in Sec.~\ref{sec:detector-limits}.

\section{Probability distributions for couplings}
\label{app:coupling-distribution}

In this appendix we include an example plot illustrating the Feldman-Cousins~\cite{Feldman:1997qc} procedure for finding the uncertainties especially useful for non-symmetric distributions. 

Figure~\ref{fig:example_1sigma_distribution} shows the distribution of the calculated 90\% C.L. upper limits on the coupling coefficient of operator $\mathcal{O}_{13}$ for protons calculated for $N_\text{trials} = 10^4$ MC generated one-body density matrix files with the procedure described in Sec.~\ref{sec:Tools-and-Methods}. The obtained distribution of the coupling coefficients is highly non-normally distributed although each of the one-body density matrix elements distributions were constructed with Gaussians. 

\section{Tests of nuclear models}
\label{app:nuclear-tests}

\begin{table}
\begin{tabular}{|c|c|c|c|}
\hline
level $J^\pi_n$ & expt. & JJ55 & GCN \\
\hline \hline
\multicolumn{4}{|c|}{$^{128}$Xe~\cite{ELEKES2015191}} \\
\hline
$2_1^+$ & 0.443 & 0.483 & 0.542 \\
\hline
$2_2^+$ & 0.969 & 1.039 & 1.259 \\
\hline
$4_1^+$ & 1.033 & 1.074 & 1.120 \\
\hline
\multicolumn{4}{|c|}{$^{129}$Xe~\cite{TIMAR2014143}} \\
\hline
$3/2_1^+$ & 0.040 & 0.133 & 0.117 \\
\hline
$3/2_2^+$ & 0.318 & 0.314 & 0.378 \\
\hline
$5/2_1^+$ & 0.321 & 0.366 & 0.406 \\
\hline
\multicolumn{4}{|c|}{$^{130}$Xe~\cite{SINGH200133}} \\
\hline
$2_1^+$ & 0.536 & 0.596 & 0.619 \\
\hline
$2_2^+$ & 1.112 & 1.219 & 1.322 \\
\hline
$4_1^+$ & 1.2043 & 1.286 & 1.311 \\
\hline
\multicolumn{4}{|c|}{$^{131}$Xe~\cite{KHAZOV20062715}} \\
\hline
$1/2_1^+$ & 0.080 & -0.078 & 0.037 \\
\hline
$5/2_1^+$ & 0.364 & 0.260 & 0.396 \\
\hline
$3/2_2^+$ & 0.405 & 0.318 & 0.471 \\
\hline
\multicolumn{4}{|c|}{$^{132}$Xe~\cite{yu2005nuclear}} \\
\hline
$2_1^+$ & 0.668 & 0.726 & 0.751 \\
\hline
$2_2^+$ & 1.298 & 1.385 & 1.425 \\
\hline
$4_1^+$ & 1.440 & 1.491 & 1.551 \\
\hline
\multicolumn{4}{|c|}{$^{134}$Xe~\cite{SONZOGNI20041}} \\
\hline
$2_1^+$ & 0.847 & 0.915 & 0.886 \\
\hline
$2_2^+$ & 1.614 & 1.719 & 1.623 \\
\hline
$0_2^+$ & 1.636 & 1.539 & 1.825 \\
\hline
$4_1^+$ & 1.731 & 1.771 & 1.775 \\
\hline
\multicolumn{4}{|c|}{$^{136}$Xe~\cite{MCCUTCHAN2018331} }\\
\hline
$2_1^+$ & $1.313$ & 1.329 & 1.363 \\
 \hline
  $4_1^+$  & $1.694 $ & 1.660 &  1.747  \\
\hline
\end{tabular}

\caption{Excitation energies in Xe isotopes, comparing experiment 
against two models, JJ55~\cite{Brown:2004xk}  and GCN5082~\cite{Caurier:2010az}. All energies are in MeV. 
Here the subscript $n$ denotes the order in the spectrum for a given $J^\pi$:
thus $2_1^+$ is the lowest $2^+$ level and $2_2^+$ is the next lowest.
All Xe isotopes with even $A$ have an empirical $J^\pi=0^+$ ground state, 
while $^{129}$Xe has a $1/2^+$ ground state and $^{131}$Xe has a 
$3/2^+$ ground state. Note that the JJ55 model puts the $1/2_1^+$ 
level below the $3/2_1^+$ level, at odds with experiment, but the 
discrepancy is within the typical theory uncertainty, about 126-150 keV, 
of similar configuration-interaction calculations~\cite{PhysRevC.74.034315,fox_usdb}. 
The experimental errors on the energies are sub-keV.
\label{tab:xe-energies}}
\end{table}

Our Monte Carlo sampling is built upon 
two input models, 
described in Section~\ref{sec:Nuclear-response-functions}.  
Both models have been previously compared against experimental 
data. The original JJ55 paper~\cite{Brown:2004xk} found good agreement with 
 excitation spectra of selected Sn, Sb, and Te isotopes as well as 
magnetic dipole (M1) moments of selected states of Sn, Te, Xe, and 
Ba isotopes. The original GCN5082 papers~\cite{Caurier:2010az} found good agreement with experimental spectra of light Xe isotopes. 
Several subsequent papers compared properties using both models: 
spectrum, transition strengths, and $g$ factors for 
$^{130}$Te~\cite{PhysRevC.105.024329}; and the energy spectra (levels) of $^{130}$Ba~\cite{PhysRevC.104.034309} and of 
$^{131}$Xe~\cite{PhysRevC.98.014309}.
Using JJ55, comparisons to experiment have been 
made of electric quadrupole (E2) and magnetic dipole (M1) transitions in Te isotopes~\cite{PhysRevC.106.034306,PhysRevC.105.034319} and in $^{131}$I and $^{132}$Xe~\cite{PhysRevC.99.014306}.
Finally, we note that GCN5082 has been used previously for calculations of dark matter cross sections~\cite{PhysRevD.88.083516}.

Here we provide additional comparison of  
observables of those two models to 
experimental results.  Specifically we compare against the measured low-lying excitation 
spectra, electric quadrupole gamma decays, and static magnetic dipole moments.  Following 
common practice for valence shell calculations~\cite{BG77}, we use effective charges and $g$-factors, with values taken from the literature, and do not use any explicit current corrections.

Table~\ref{tab:xe-energies} compares experimental excitation energies of 
low-lying states against the values calculated in our two models. 
Note that the typical theory error in shell model calculations is of the 
order of 126 keV~\cite{PhysRevC.74.034315} to 150 keV~\cite{fox_usdb}.

\begin{table}
\begin{tabular}{|c|c|c|c|c|c|}
\hline
Transition & $E_\gamma$ & $t_{1/2}$ & B(E2) &  JJ55 & GCN \\
  & (MeV) & ps & $e^2$-fm$^4$ & $e^2$-fm$^4$ & $e^2$-fm$^4$ \\
  \hline \hline 
 \multicolumn{6}{|c|}{$^{128}$Xe~\cite{ELEKES2015191}} \\ 
 \hline
 $2_1^+ \rightarrow 0^+_1$ & 0.443 & $18 \pm 4$  &  $1845 \pm 410$ & 1670 & 1560 \\
 \hline
  $4_1^+ \rightarrow 2^+_1$ & 0.590 & $3.33 \pm 0.14 $ & $2380 \pm 100 $ & 2450 & 2360 \\
\hline
\multicolumn{6}{|c|}{$^{130}$Xe~\cite{SINGH200133}} \\
\hline
$2_1^+ \rightarrow 0^+_1$ & 0.536 & $8.6 \pm 1.5 $ &  $1490 \pm 260 $ & 1340  &  1330 \\
 \hline
\multicolumn{6}{|c|}{$^{132}$Xe~\cite{yu2005nuclear}} \\
 \hline
 $2_1^+ \rightarrow 0^+_1$ & 0.668 & $4.63 \pm 0.30$ & $920 \pm 60 $  &  990  & 1010   \\
 \hline
  $4_1^+ \rightarrow 2^+_1$ & 0.773 & $1.80 \pm 0.14$ & $1140 \pm 90 $ & 1460 &  1520  \\
\hline
\multicolumn{6}{|c|}{$^{134}$Xe~\cite{SONZOGNI20041}} \\
 \hline
 $2_1^+ \rightarrow 0^+_1$ & 0.847 & $2.08 \pm 0 .14$ & $620 \pm 40$  & 640  &  610  \\
 \hline
  $4_1^+ \rightarrow 2^+_1$ & 0.884 & $2.22 \pm 0.14$ & $480 \pm 30 $ &  390  &  540  \\
\hline
\multicolumn{6}{|c|}{$^{136}$Xe~\cite{MCCUTCHAN2018331} }\\
 \hline
 $2_1^+ \rightarrow 0^+_1$ & 1.313 & $0.360 \pm 0.014$ & $400 \pm 16$  &  470  &  520   \\
 \hline
  $4_1^+ \rightarrow 2^+_1$ & 0.381 & $1290 \pm 17 $ & $55 \pm 1 $ & 56 &  180  \\
\hline
\end{tabular}
\caption{\label{tab:be2} 
Reduced electric quadrupole transition probabilities, B(E2)s, comparing experiment 
(``B(E2)'') against values using the JJ55 and 
GCN models.}
\end{table}

In Table \ref{tab:be2} we compare experimental reduced 
electric quadrupole matrix elements, or B(E2)s, 
against calculated values. For our calculations, we used the standard 
prescriptions~\cite{BG77,Suhonen2007},
working in a single-particle harmonic oscillator basis with an oscillator length parameter of 2.295 fm, and adopted the 
effective charges of 1.86 $e$ and 0.65 $e$ for 
valence protons and neutrons, respectively~\cite{PhysRevC.106.034306}. We only considered even mass numbers $A$  to avoid the complication of mixing E2 and M1 transitions.

\begin{table}
\begin{tabular}{|c|c|c|c|}
\hline
 & \multicolumn{3}{c|} {$\mu/\mu_N$ }\\

level & expt & JJ55 & GCN 
   \\
  \hline \hline 
 \multicolumn{4}{|c|}{$^{128}$Xe~\cite{PhysRevC.12.628} } \\ 
 \hline
 $2_1^+ $ &   $+0.68 \pm 0.07$ & +0.610 &  +0.809 \\
\hline
 \multicolumn{4}{|c|}{$^{129}$Xe~\cite{Brinkmann62,van1974study} } \\ 
 \hline
 $1/2_1^+ $ &   $-0.777961(16) $ & -0.823 & -0.889 \\
\hline
 $3/2_1^+ $ &   $+0.58 \pm 0.08$ & +0.638 &  +0.588 \\
\hline
\multicolumn{4}{|c|}{$^{130}$Xe~\cite{PhysRevC.65.024316}} \\
\hline
$2_1^+ $  &  $+0.67\pm 0.10$ &  +0.580 &  +0.776 \\
 \hline
 $2_2^+ $  &  $+0.9\pm 0.2$ & +0.585  &  +0.72 \\
 \hline
$4_1^+ $  &  $+1.7\pm 0.2$ &  +1.57  &  +1.83 \\
 \hline 
 \multicolumn{4}{|c|}{$^{131}$Xe~\cite{Brinkmann62}} \\
\hline
$3/2_1^+ $  &  $+0.691862(4)$ & +0.753 &  +0.816 \\
 \hline
\multicolumn{4}{|c|}{$^{132}$Xe~\cite{PhysRevC.65.024316}} \\
 \hline
 $2_1^+ $ &   $+0.63\pm 0.10$ &  +0.56  & +0.70   \\
 \hline
  $2_2^+ $ &   $+0.2\pm 0.4$  &  +0.32  & +0.50   \\
 \hline
  $4_1^+ $  & $+2.4\pm 0.6$ & +1.35$^*$ &  +2.17  \\
\hline
\multicolumn{4}{|c|}{$^{134}$Xe~\cite{PhysRevC.65.024316} } \\
 \hline
 $2_1^+ $  & $ +0.708 \pm 0.014$  & +0.688  &  +0.559  \\
 \hline
  $4_1^+ $  & $+3.2 \pm 0.6$ &   +2.88  & +2.980 \\
\hline
\multicolumn{4}{|c|}{$^{136}$Xe~\cite{PhysRevC.65.024316, PhysRevC.31.570} }\\
 \hline
$2_1^+$ & $+1.53 \pm 0.24$ &  +1.52 & +1.53 \\
 \hline
  $4_1^+$  & $+3.2 \pm 0.6$ & +2.94 &  +3.10  \\
\hline
\end{tabular}
\caption{\label{tab:m1moment} 
Magnetic dipole moments, in units of the nuclear magneton $\mu_N$, comparing experiment  against values using the JJ55 and 
GCN models.  Note: for the JJ55 calculation of $^{132}$Xe, the $4_2^+$ level has a moment 
of +2.19 $\mu_N$, so it is possible the calculated $4_{1,2}^+$ levels are inverted.}
\end{table}

Finally, in Table~\ref{tab:m1moment}, we compare experimental static magnetic dipole moments, or $\mu$, against calculated values. Following \cite{Brown:2004xk}, we quench the free spin $g$ factors by 0.7. 

Broadly speaking, both models give similar agreement with experiment, 
justifying using them as an equally weighted basis for our 
calculations.

\newpage
\phantom{i}
\newpage

\bibliography{dmscatter-uncertainties}

\providecommand{\noopsort}[1]{}\providecommand{\singleletter}[1]{#1}%
\begin{thebibliography}{72}%
\makeatletter
\providecommand \@ifxundefined [1]{%
 \@ifx{#1\undefined}
}%
\providecommand \@ifnum [1]{%
 \ifnum #1\expandafter \@firstoftwo
 \else \expandafter \@secondoftwo
 \fi
}%
\providecommand \@ifx [1]{%
 \ifx #1\expandafter \@firstoftwo
 \else \expandafter \@secondoftwo
 \fi
}%
\providecommand \natexlab [1]{#1}%
\providecommand \enquote  [1]{``#1''}%
\providecommand \bibnamefont  [1]{#1}%
\providecommand \bibfnamefont [1]{#1}%
\providecommand \citenamefont [1]{#1}%
\providecommand \href@noop [0]{\@secondoftwo}%
\providecommand \href [0]{\begingroup \@sanitize@url \@href}%
\providecommand \@href[1]{\@@startlink{#1}\@@href}%
\providecommand \@@href[1]{\endgroup#1\@@endlink}%
\providecommand \@sanitize@url [0]{\catcode `\\12\catcode `\$12\catcode
  `\&12\catcode `\#12\catcode `\^12\catcode `\_12\catcode `\%12\relax}%
\providecommand \@@startlink[1]{}%
\providecommand \@@endlink[0]{}%
\providecommand \url  [0]{\begingroup\@sanitize@url \@url }%
\providecommand \@url [1]{\endgroup\@href {#1}{\urlprefix }}%
\providecommand \urlprefix  [0]{URL }%
\providecommand \Eprint [0]{\href }%
\providecommand \doibase [0]{http://dx.doi.org/}%
\providecommand \selectlanguage [0]{\@gobble}%
\providecommand \bibinfo  [0]{\@secondoftwo}%
\providecommand \bibfield  [0]{\@secondoftwo}%
\providecommand \translation [1]{[#1]}%
\providecommand \BibitemOpen [0]{}%
\providecommand \bibitemStop [0]{}%
\providecommand \bibitemNoStop [0]{.\EOS\space}%
\providecommand \EOS [0]{\spacefactor3000\relax}%
\providecommand \BibitemShut  [1]{\csname bibitem#1\endcsname}%
\let\auto@bib@innerbib\@empty
\bibitem [{\citenamefont {Akerib}\ \emph {et~al.}(2022)\citenamefont {Akerib}
  \emph {et~al.}}]{Akerib:2022ort}%
  \BibitemOpen
  \bibfield  {author} {\bibinfo {author} {\bibfnamefont {D.~S.}\ \bibnamefont
  {Akerib}} \emph {et~al.},\ }\bibfield  {title} {\enquote {\bibinfo {title}
  {{Snowmass2021 Cosmic Frontier Dark Matter Direct Detection to the Neutrino
  Fog}},}\ }in\ \href@noop {} {\emph {\bibinfo {booktitle} {{2022 Snowmass
  Summer Study}}}}\ (\bibinfo {year} {2022})\ \Eprint
  {http://arxiv.org/abs/2203.08084} {arXiv:2203.08084 [hep-ex]} \BibitemShut
  {NoStop}%
\bibitem [{\citenamefont {Aprile}\ \emph
  {et~al.}(2022{\natexlab{a}})\citenamefont {Aprile} \emph
  {et~al.}}]{XENONCollaboration:2022kmb}%
  \BibitemOpen
  \bibfield  {author} {\bibinfo {author} {\bibfnamefont {E.}~\bibnamefont
  {Aprile}} \emph {et~al.} (\bibinfo {collaboration} {(XENON
  Collaboration)\textdagger{}\textdagger{}, XENON}),\ }\bibfield  {title}
  {\enquote {\bibinfo {title} {{Search for New Physics in Electronic Recoil
  Data from XENONnT}},}\ }\href {\doibase 10.1103/PhysRevLett.129.161805}
  {\bibfield  {journal} {\bibinfo  {journal} {Phys. Rev. Lett.}\ }\textbf
  {\bibinfo {volume} {129}},\ \bibinfo {pages} {161805} (\bibinfo {year}
  {2022}{\natexlab{a}})},\ \Eprint {http://arxiv.org/abs/2207.11330}
  {arXiv:2207.11330 [hep-ex]} \BibitemShut {NoStop}%
\bibitem [{\citenamefont {Aprile}\ \emph {et~al.}(2023)\citenamefont {Aprile}
  \emph {et~al.}}]{XENON:2023sxq}%
  \BibitemOpen
  \bibfield  {author} {\bibinfo {author} {\bibfnamefont {E.}~\bibnamefont
  {Aprile}} \emph {et~al.} (\bibinfo {collaboration} {XENON}),\ }\bibfield
  {title} {\enquote {\bibinfo {title} {{First Dark Matter Search with Nuclear
  Recoils from the XENONnT Experiment}},}\ }\href@noop {} {\  (\bibinfo {year}
  {2023})},\ \Eprint {http://arxiv.org/abs/2303.14729} {arXiv:2303.14729
  [hep-ex]} \BibitemShut {NoStop}%
\bibitem [{\citenamefont {Meng}\ \emph {et~al.}(2021)\citenamefont {Meng} \emph
  {et~al.}}]{PandaX-4T:2021bab}%
  \BibitemOpen
  \bibfield  {author} {\bibinfo {author} {\bibfnamefont {Yue}\ \bibnamefont
  {Meng}} \emph {et~al.} (\bibinfo {collaboration} {PandaX-4T}),\ }\bibfield
  {title} {\enquote {\bibinfo {title} {{Dark Matter Search Results from the
  PandaX-4T Commissioning Run}},}\ }\href {\doibase
  10.1103/PhysRevLett.127.261802} {\bibfield  {journal} {\bibinfo  {journal}
  {Phys. Rev. Lett.}\ }\textbf {\bibinfo {volume} {127}},\ \bibinfo {pages}
  {261802} (\bibinfo {year} {2021})},\ \Eprint
  {http://arxiv.org/abs/2107.13438} {arXiv:2107.13438 [hep-ex]} \BibitemShut
  {NoStop}%
\bibitem [{\citenamefont {Roszkowski}\ \emph {et~al.}(2018)\citenamefont
  {Roszkowski}, \citenamefont {Sessolo},\ and\ \citenamefont
  {Trojanowski}}]{Roszkowski:2017nbc}%
  \BibitemOpen
  \bibfield  {author} {\bibinfo {author} {\bibfnamefont {Leszek}\ \bibnamefont
  {Roszkowski}}, \bibinfo {author} {\bibfnamefont {Enrico~Maria}\ \bibnamefont
  {Sessolo}}, \ and\ \bibinfo {author} {\bibfnamefont {Sebastian}\ \bibnamefont
  {Trojanowski}},\ }\bibfield  {title} {\enquote {\bibinfo {title} {{WIMP dark
  matter candidates and searches\textemdash{}current status and future
  prospects}},}\ }\href {\doibase 10.1088/1361-6633/aab913} {\bibfield
  {journal} {\bibinfo  {journal} {Rept. Prog. Phys.}\ }\textbf {\bibinfo
  {volume} {81}},\ \bibinfo {pages} {066201} (\bibinfo {year} {2018})},\
  \Eprint {http://arxiv.org/abs/1707.06277} {arXiv:1707.06277 [hep-ph]}
  \BibitemShut {NoStop}%
\bibitem [{\citenamefont {Akerib}\ \emph {et~al.}(2017)\citenamefont {Akerib}
  \emph {et~al.}}]{LUX:2017ree}%
  \BibitemOpen
  \bibfield  {author} {\bibinfo {author} {\bibfnamefont {D.~S.}\ \bibnamefont
  {Akerib}} \emph {et~al.} (\bibinfo {collaboration} {LUX}),\ }\bibfield
  {title} {\enquote {\bibinfo {title} {{Limits on spin-dependent WIMP-nucleon
  cross section obtained from the complete LUX exposure}},}\ }\href {\doibase
  10.1103/PhysRevLett.118.251302} {\bibfield  {journal} {\bibinfo  {journal}
  {Phys. Rev. Lett.}\ }\textbf {\bibinfo {volume} {118}},\ \bibinfo {pages}
  {251302} (\bibinfo {year} {2017})},\ \Eprint
  {http://arxiv.org/abs/1705.03380} {arXiv:1705.03380 [astro-ph.CO]}
  \BibitemShut {NoStop}%
\bibitem [{\citenamefont {Aprile}\ \emph {et~al.}(2018)\citenamefont {Aprile}
  \emph {et~al.}}]{XENON:2018voc}%
  \BibitemOpen
  \bibfield  {author} {\bibinfo {author} {\bibfnamefont {E.}~\bibnamefont
  {Aprile}} \emph {et~al.} (\bibinfo {collaboration} {XENON}),\ }\bibfield
  {title} {\enquote {\bibinfo {title} {{Dark Matter Search Results from a One
  Ton-Year Exposure of XENON1T}},}\ }\href {\doibase
  10.1103/PhysRevLett.121.111302} {\bibfield  {journal} {\bibinfo  {journal}
  {Phys. Rev. Lett.}\ }\textbf {\bibinfo {volume} {121}},\ \bibinfo {pages}
  {111302} (\bibinfo {year} {2018})},\ \Eprint
  {http://arxiv.org/abs/1805.12562} {arXiv:1805.12562 [astro-ph.CO]}
  \BibitemShut {NoStop}%
\bibitem [{\citenamefont {Agnes}\ \emph {et~al.}(2018)\citenamefont {Agnes}
  \emph {et~al.}}]{DarkSide:2018bpj}%
  \BibitemOpen
  \bibfield  {author} {\bibinfo {author} {\bibfnamefont {P.}~\bibnamefont
  {Agnes}} \emph {et~al.} (\bibinfo {collaboration} {DarkSide}),\ }\bibfield
  {title} {\enquote {\bibinfo {title} {{Low-Mass Dark Matter Search with the
  DarkSide-50 Experiment}},}\ }\href {\doibase 10.1103/PhysRevLett.121.081307}
  {\bibfield  {journal} {\bibinfo  {journal} {Phys. Rev. Lett.}\ }\textbf
  {\bibinfo {volume} {121}},\ \bibinfo {pages} {081307} (\bibinfo {year}
  {2018})},\ \Eprint {http://arxiv.org/abs/1802.06994} {arXiv:1802.06994
  [astro-ph.HE]} \BibitemShut {NoStop}%
\bibitem [{\citenamefont {Aalbers}\ \emph {et~al.}(2022)\citenamefont {Aalbers}
  \emph {et~al.}}]{LZ:2022ufs}%
  \BibitemOpen
  \bibfield  {author} {\bibinfo {author} {\bibfnamefont {J.}~\bibnamefont
  {Aalbers}} \emph {et~al.} (\bibinfo {collaboration} {LZ}),\ }\bibfield
  {title} {\enquote {\bibinfo {title} {{First Dark Matter Search Results from
  the LUX-ZEPLIN (LZ) Experiment}},}\ }\href@noop {} {\  (\bibinfo {year}
  {2022})},\ \Eprint {http://arxiv.org/abs/2207.03764} {arXiv:2207.03764
  [hep-ex]} \BibitemShut {NoStop}%
\bibitem [{\citenamefont {Fitzpatrick}\ \emph {et~al.}(2013)\citenamefont
  {Fitzpatrick}, \citenamefont {Haxton}, \citenamefont {Katz}, \citenamefont
  {Lubbers},\ and\ \citenamefont {Xu}}]{Fitzpatrick:2012ix}%
  \BibitemOpen
  \bibfield  {author} {\bibinfo {author} {\bibfnamefont {A.~Liam}\ \bibnamefont
  {Fitzpatrick}}, \bibinfo {author} {\bibfnamefont {Wick}\ \bibnamefont
  {Haxton}}, \bibinfo {author} {\bibfnamefont {Emanuel}\ \bibnamefont {Katz}},
  \bibinfo {author} {\bibfnamefont {Nicholas}\ \bibnamefont {Lubbers}}, \ and\
  \bibinfo {author} {\bibfnamefont {Yiming}\ \bibnamefont {Xu}},\ }\bibfield
  {title} {\enquote {\bibinfo {title} {{The Effective Field Theory of Dark
  Matter Direct Detection}},}\ }\href {\doibase 10.1088/1475-7516/2013/02/004}
  {\bibfield  {journal} {\bibinfo  {journal} {JCAP}\ }\textbf {\bibinfo
  {volume} {02}},\ \bibinfo {pages} {004} (\bibinfo {year} {2013})},\ \Eprint
  {http://arxiv.org/abs/1203.3542} {arXiv:1203.3542 [hep-ph]} \BibitemShut
  {NoStop}%
\bibitem [{\citenamefont {Anand}\ \emph {et~al.}(2014)\citenamefont {Anand},
  \citenamefont {Fitzpatrick},\ and\ \citenamefont {Haxton}}]{Anand:2013yka}%
  \BibitemOpen
  \bibfield  {author} {\bibinfo {author} {\bibfnamefont {Nikhil}\ \bibnamefont
  {Anand}}, \bibinfo {author} {\bibfnamefont {A.~Liam}\ \bibnamefont
  {Fitzpatrick}}, \ and\ \bibinfo {author} {\bibfnamefont {W.~C.}\ \bibnamefont
  {Haxton}},\ }\bibfield  {title} {\enquote {\bibinfo {title} {{Weakly
  interacting massive particle-nucleus elastic scattering response}},}\ }\href
  {\doibase 10.1103/PhysRevC.89.065501} {\bibfield  {journal} {\bibinfo
  {journal} {Phys. Rev. C}\ }\textbf {\bibinfo {volume} {89}},\ \bibinfo
  {pages} {065501} (\bibinfo {year} {2014})},\ \Eprint
  {http://arxiv.org/abs/1308.6288} {arXiv:1308.6288 [hep-ph]} \BibitemShut
  {NoStop}%
\bibitem [{\citenamefont {Agnes}\ \emph {et~al.}(2020)\citenamefont {Agnes}
  \emph {et~al.}}]{DarkSide-50:2020swd}%
  \BibitemOpen
  \bibfield  {author} {\bibinfo {author} {\bibfnamefont {P.}~\bibnamefont
  {Agnes}} \emph {et~al.} (\bibinfo {collaboration} {DarkSide-50}),\ }\bibfield
   {title} {\enquote {\bibinfo {title} {{Effective field theory interactions
  for liquid argon target in DarkSide-50 experiment}},}\ }\href {\doibase
  10.1103/PhysRevD.101.062002} {\bibfield  {journal} {\bibinfo  {journal}
  {Phys. Rev. D}\ }\textbf {\bibinfo {volume} {101}},\ \bibinfo {pages}
  {062002} (\bibinfo {year} {2020})},\ \Eprint
  {http://arxiv.org/abs/2002.07794} {arXiv:2002.07794 [hep-ex]} \BibitemShut
  {NoStop}%
\bibitem [{\citenamefont {Akerib}\ \emph {et~al.}(2021)\citenamefont {Akerib}
  \emph {et~al.}}]{LUX:2021ksq}%
  \BibitemOpen
  \bibfield  {author} {\bibinfo {author} {\bibfnamefont {D.~S.}\ \bibnamefont
  {Akerib}} \emph {et~al.} (\bibinfo {collaboration} {LUX}),\ }\bibfield
  {title} {\enquote {\bibinfo {title} {{Constraints on effective field theory
  couplings using 311.2 days of LUX data}},}\ }\href {\doibase
  10.1103/PhysRevD.104.062005} {\bibfield  {journal} {\bibinfo  {journal}
  {Phys. Rev. D}\ }\textbf {\bibinfo {volume} {104}},\ \bibinfo {pages}
  {062005} (\bibinfo {year} {2021})},\ \Eprint
  {http://arxiv.org/abs/2102.06998} {arXiv:2102.06998 [astro-ph.CO]}
  \BibitemShut {NoStop}%
\bibitem [{\citenamefont {Aprile}\ \emph
  {et~al.}(2022{\natexlab{b}})\citenamefont {Aprile} \emph
  {et~al.}}]{XENON:2022avm}%
  \BibitemOpen
  \bibfield  {author} {\bibinfo {author} {\bibfnamefont {E.}~\bibnamefont
  {Aprile}} \emph {et~al.} (\bibinfo {collaboration} {XENON}),\ }\bibfield
  {title} {\enquote {\bibinfo {title} {{Effective Field Theory and Inelastic
  Dark Matter Results from XENON1T}},}\ }\href@noop {} {\  (\bibinfo {year}
  {2022}{\natexlab{b}})},\ \Eprint {http://arxiv.org/abs/2210.07591}
  {arXiv:2210.07591 [hep-ex]} \BibitemShut {NoStop}%
\bibitem [{\citenamefont {Cirigliano}\ \emph {et~al.}(2012)\citenamefont
  {Cirigliano}, \citenamefont {Graesser},\ and\ \citenamefont
  {Ovanesyan}}]{Cirigliano:2012pq}%
  \BibitemOpen
  \bibfield  {author} {\bibinfo {author} {\bibfnamefont {Vincenzo}\
  \bibnamefont {Cirigliano}}, \bibinfo {author} {\bibfnamefont {Michael~L.}\
  \bibnamefont {Graesser}}, \ and\ \bibinfo {author} {\bibfnamefont {Grigory}\
  \bibnamefont {Ovanesyan}},\ }\bibfield  {title} {\enquote {\bibinfo {title}
  {{WIMP-nucleus scattering in chiral effective theory}},}\ }\href {\doibase
  10.1007/JHEP10(2012)025} {\bibfield  {journal} {\bibinfo  {journal} {JHEP}\
  }\textbf {\bibinfo {volume} {10}},\ \bibinfo {pages} {025} (\bibinfo {year}
  {2012})},\ \Eprint {http://arxiv.org/abs/1205.2695} {arXiv:1205.2695
  [hep-ph]} \BibitemShut {NoStop}%
\bibitem [{\citenamefont {Hoferichter}\ \emph {et~al.}(2015)\citenamefont
  {Hoferichter}, \citenamefont {Klos},\ and\ \citenamefont
  {Schwenk}}]{hoferichter2015chiral}%
  \BibitemOpen
  \bibfield  {author} {\bibinfo {author} {\bibfnamefont {Martin}\ \bibnamefont
  {Hoferichter}}, \bibinfo {author} {\bibfnamefont {Philipp}\ \bibnamefont
  {Klos}}, \ and\ \bibinfo {author} {\bibfnamefont {Achim}\ \bibnamefont
  {Schwenk}},\ }\bibfield  {title} {\enquote {\bibinfo {title} {Chiral power
  counting of one-and two-body currents in direct detection of dark matter},}\
  }\href@noop {} {\bibfield  {journal} {\bibinfo  {journal} {Physics Letters
  B}\ }\textbf {\bibinfo {volume} {746}},\ \bibinfo {pages} {410--416}
  (\bibinfo {year} {2015})}\BibitemShut {NoStop}%
\bibitem [{\citenamefont {Bishara}\ \emph
  {et~al.}(2017{\natexlab{a}})\citenamefont {Bishara}, \citenamefont {Brod},
  \citenamefont {Grinstein},\ and\ \citenamefont {Zupan}}]{Bishara_2017}%
  \BibitemOpen
  \bibfield  {author} {\bibinfo {author} {\bibfnamefont {Fady}\ \bibnamefont
  {Bishara}}, \bibinfo {author} {\bibfnamefont {Joachim}\ \bibnamefont {Brod}},
  \bibinfo {author} {\bibfnamefont {Benjamin}\ \bibnamefont {Grinstein}}, \
  and\ \bibinfo {author} {\bibfnamefont {Jure}\ \bibnamefont {Zupan}},\
  }\bibfield  {title} {\enquote {\bibinfo {title} {Chiral effective theory of
  dark matter direct detection},}\ }\href {\doibase
  10.1088/1475-7516/2017/02/009} {\bibfield  {journal} {\bibinfo  {journal}
  {Journal of Cosmology and Astroparticle Physics}\ }\textbf {\bibinfo {volume}
  {2017}},\ \bibinfo {pages} {009} (\bibinfo {year}
  {2017}{\natexlab{a}})}\BibitemShut {NoStop}%
\bibitem [{\citenamefont {Brod}\ \emph {et~al.}(2018)\citenamefont {Brod},
  \citenamefont {Gootjes-Dreesbach}, \citenamefont {Tammaro},\ and\
  \citenamefont {Zupan}}]{Brod:2017bsw}%
  \BibitemOpen
  \bibfield  {author} {\bibinfo {author} {\bibfnamefont {Joachim}\ \bibnamefont
  {Brod}}, \bibinfo {author} {\bibfnamefont {Aaron}\ \bibnamefont
  {Gootjes-Dreesbach}}, \bibinfo {author} {\bibfnamefont {Michele}\
  \bibnamefont {Tammaro}}, \ and\ \bibinfo {author} {\bibfnamefont {Jure}\
  \bibnamefont {Zupan}},\ }\bibfield  {title} {\enquote {\bibinfo {title}
  {{Effective Field Theory for Dark Matter Direct Detection up to Dimension
  Seven}},}\ }\href {\doibase 10.1007/JHEP10(2018)065} {\bibfield  {journal}
  {\bibinfo  {journal} {JHEP}\ }\textbf {\bibinfo {volume} {10}},\ \bibinfo
  {pages} {065} (\bibinfo {year} {2018})},\ \bibinfo {note} {[Erratum: JHEP 07,
  012 (2023)]},\ \Eprint {http://arxiv.org/abs/1710.10218} {arXiv:1710.10218
  [hep-ph]} \BibitemShut {NoStop}%
\bibitem [{\citenamefont {Hoferichter}\ \emph
  {et~al.}(2019{\natexlab{a}})\citenamefont {Hoferichter}, \citenamefont
  {Klos}, \citenamefont {Menéndez},\ and\ \citenamefont
  {Schwenk}}]{hoferichter2019darkmatternucleus}%
  \BibitemOpen
  \bibfield  {author} {\bibinfo {author} {\bibfnamefont {Martin}\ \bibnamefont
  {Hoferichter}}, \bibinfo {author} {\bibfnamefont {Philipp}\ \bibnamefont
  {Klos}}, \bibinfo {author} {\bibfnamefont {Javier}\ \bibnamefont
  {Menéndez}}, \ and\ \bibinfo {author} {\bibfnamefont {Achim}\ \bibnamefont
  {Schwenk}},\ }\href@noop {} {\enquote {\bibinfo {title} {Dark-matter-nucleus
  scattering in chiral effective field theory},}\ } (\bibinfo {year}
  {2019}{\natexlab{a}}),\ \Eprint {http://arxiv.org/abs/1903.11075}
  {arXiv:1903.11075 [hep-ph]} \BibitemShut {NoStop}%
\bibitem [{\citenamefont {Hoferichter}\ \emph
  {et~al.}(2019{\natexlab{b}})\citenamefont {Hoferichter}, \citenamefont
  {Klos}, \citenamefont {Men\'endez},\ and\ \citenamefont
  {Schwenk}}]{PhysRevD.99.055031}%
  \BibitemOpen
  \bibfield  {author} {\bibinfo {author} {\bibfnamefont {Martin}\ \bibnamefont
  {Hoferichter}}, \bibinfo {author} {\bibfnamefont {Philipp}\ \bibnamefont
  {Klos}}, \bibinfo {author} {\bibfnamefont {Javier}\ \bibnamefont
  {Men\'endez}}, \ and\ \bibinfo {author} {\bibfnamefont {Achim}\ \bibnamefont
  {Schwenk}},\ }\bibfield  {title} {\enquote {\bibinfo {title} {Nuclear
  structure factors for general spin-independent wimp-nucleus scattering},}\
  }\href {\doibase 10.1103/PhysRevD.99.055031} {\bibfield  {journal} {\bibinfo
  {journal} {Phys. Rev. D}\ }\textbf {\bibinfo {volume} {99}},\ \bibinfo
  {pages} {055031} (\bibinfo {year} {2019}{\natexlab{b}})}\BibitemShut
  {NoStop}%
\bibitem [{\citenamefont {Richardson}\ \emph {et~al.}(2022)\citenamefont
  {Richardson}, \citenamefont {Lin},\ and\ \citenamefont
  {Nguyen}}]{PhysRevC.106.044003}%
  \BibitemOpen
  \bibfield  {author} {\bibinfo {author} {\bibfnamefont {Thomas~R.}\
  \bibnamefont {Richardson}}, \bibinfo {author} {\bibfnamefont {Xincheng}\
  \bibnamefont {Lin}}, \ and\ \bibinfo {author} {\bibfnamefont {Son~T.}\
  \bibnamefont {Nguyen}},\ }\bibfield  {title} {\enquote {\bibinfo {title}
  {Large-${N}_{c}$ constraints for elastic dark-matter--light-nucleus
  scattering in pionless effective field theory},}\ }\href {\doibase
  10.1103/PhysRevC.106.044003} {\bibfield  {journal} {\bibinfo  {journal}
  {Phys. Rev. C}\ }\textbf {\bibinfo {volume} {106}},\ \bibinfo {pages}
  {044003} (\bibinfo {year} {2022})}\BibitemShut {NoStop}%
\bibitem [{\citenamefont {Trickle}\ \emph {et~al.}(2022)\citenamefont
  {Trickle}, \citenamefont {Zhang},\ and\ \citenamefont
  {Zurek}}]{Trickle:2020oki}%
  \BibitemOpen
  \bibfield  {author} {\bibinfo {author} {\bibfnamefont {Tanner}\ \bibnamefont
  {Trickle}}, \bibinfo {author} {\bibfnamefont {Zhengkang}\ \bibnamefont
  {Zhang}}, \ and\ \bibinfo {author} {\bibfnamefont {Kathryn~M.}\ \bibnamefont
  {Zurek}},\ }\bibfield  {title} {\enquote {\bibinfo {title} {{Effective field
  theory of dark matter direct detection with collective excitations}},}\
  }\href {\doibase 10.1103/PhysRevD.105.015001} {\bibfield  {journal} {\bibinfo
   {journal} {Phys. Rev. D}\ }\textbf {\bibinfo {volume} {105}},\ \bibinfo
  {pages} {015001} (\bibinfo {year} {2022})},\ \Eprint
  {http://arxiv.org/abs/2009.13534} {arXiv:2009.13534 [hep-ph]} \BibitemShut
  {NoStop}%
\bibitem [{\citenamefont {Adams}\ \emph {et~al.}(2022)\citenamefont {Adams}
  \emph {et~al.}}]{ADAMS2022103902}%
  \BibitemOpen
  \bibfield  {author} {\bibinfo {author} {\bibfnamefont {D.Q.}\ \bibnamefont
  {Adams}} \emph {et~al.} (\bibinfo {collaboration} {CUORE}),\ }\bibfield
  {title} {\enquote {\bibinfo {title} {Cuore opens the door to tonne-scale
  cryogenics experiments},}\ }\href {\doibase
  https://doi.org/10.1016/j.ppnp.2021.103902} {\bibfield  {journal} {\bibinfo
  {journal} {Progress in Particle and Nuclear Physics}\ }\textbf {\bibinfo
  {volume} {122}},\ \bibinfo {pages} {103902} (\bibinfo {year}
  {2022})}\BibitemShut {NoStop}%
\bibitem [{\citenamefont {Suhonen}(2007)}]{Suhonen2007}%
  \BibitemOpen
  \bibfield  {author} {\bibinfo {author} {\bibfnamefont {Jouni}\ \bibnamefont
  {Suhonen}},\ }\enquote {\bibinfo {title} {From nucleons to nucleus: Concepts
  of microscopic nuclear theory},}\ \ (\bibinfo  {publisher} {Springer Berlin
  Heidelberg},\ \bibinfo {address} {Berlin, Heidelberg},\ \bibinfo {year}
  {2007})\BibitemShut {NoStop}%
\bibitem [{\citenamefont {Hergert}(2020)}]{Hergert:2020bxy}%
  \BibitemOpen
  \bibfield  {author} {\bibinfo {author} {\bibfnamefont {H.}~\bibnamefont
  {Hergert}},\ }\bibfield  {title} {\enquote {\bibinfo {title} {{A Guided Tour
  of $ab$ $initio$ Nuclear Many-Body Theory}},}\ }\href {\doibase
  10.3389/fphy.2020.00379} {\bibfield  {journal} {\bibinfo  {journal} {Front.
  in Phys.}\ }\textbf {\bibinfo {volume} {8}},\ \bibinfo {pages} {379}
  (\bibinfo {year} {2020})},\ \Eprint {http://arxiv.org/abs/2008.05061}
  {arXiv:2008.05061 [nucl-th]} \BibitemShut {NoStop}%
\bibitem [{\citenamefont {Hu}\ \emph {et~al.}(2022)\citenamefont {Hu},
  \citenamefont {Padua-Arg\"uelles}, \citenamefont {Leutheusser}, \citenamefont
  {Miyagi}, \citenamefont {Stroberg},\ and\ \citenamefont
  {Holt}}]{PhysRevLett.128.072502}%
  \BibitemOpen
  \bibfield  {author} {\bibinfo {author} {\bibfnamefont {B.~S.}\ \bibnamefont
  {Hu}}, \bibinfo {author} {\bibfnamefont {J.}~\bibnamefont
  {Padua-Arg\"uelles}}, \bibinfo {author} {\bibfnamefont {S.}~\bibnamefont
  {Leutheusser}}, \bibinfo {author} {\bibfnamefont {T.}~\bibnamefont {Miyagi}},
  \bibinfo {author} {\bibfnamefont {S.~R.}\ \bibnamefont {Stroberg}}, \ and\
  \bibinfo {author} {\bibfnamefont {J.~D.}\ \bibnamefont {Holt}},\ }\bibfield
  {title} {\enquote {\bibinfo {title} {Ab initio structure factors for
  spin-dependent dark matter direct detection},}\ }\href {\doibase
  10.1103/PhysRevLett.128.072502} {\bibfield  {journal} {\bibinfo  {journal}
  {Phys. Rev. Lett.}\ }\textbf {\bibinfo {volume} {128}},\ \bibinfo {pages}
  {072502} (\bibinfo {year} {2022})}\BibitemShut {NoStop}%
\bibitem [{\citenamefont {Gorton}\ \emph {et~al.}(2023)\citenamefont {Gorton},
  \citenamefont {Johnson}, \citenamefont {Jiao},\ and\ \citenamefont
  {Nikoleyczik}}]{Gorton:2022eed}%
  \BibitemOpen
  \bibfield  {author} {\bibinfo {author} {\bibfnamefont {Oliver~C.}\
  \bibnamefont {Gorton}}, \bibinfo {author} {\bibfnamefont {Calvin~W.}\
  \bibnamefont {Johnson}}, \bibinfo {author} {\bibfnamefont {Changfeng}\
  \bibnamefont {Jiao}}, \ and\ \bibinfo {author} {\bibfnamefont {Jonathan}\
  \bibnamefont {Nikoleyczik}},\ }\bibfield  {title} {\enquote {\bibinfo {title}
  {{dmscatter: A fast program for WIMP-nucleus scattering}},}\ }\href {\doibase
  10.1016/j.cpc.2022.108597} {\bibfield  {journal} {\bibinfo  {journal}
  {Comput. Phys. Commun.}\ }\textbf {\bibinfo {volume} {284}},\ \bibinfo
  {pages} {108597} (\bibinfo {year} {2023})},\ \Eprint
  {http://arxiv.org/abs/2209.09187} {arXiv:2209.09187 [nucl-th]} \BibitemShut
  {NoStop}%
\bibitem [{\citenamefont {Donnelly}\ and\ \citenamefont
  {Haxton}(1979)}]{Donnelly:1979ezn}%
  \BibitemOpen
  \bibfield  {author} {\bibinfo {author} {\bibfnamefont {T.~W.}\ \bibnamefont
  {Donnelly}}\ and\ \bibinfo {author} {\bibfnamefont {W.~C.}\ \bibnamefont
  {Haxton}},\ }\bibfield  {title} {\enquote {\bibinfo {title} {{Multipole
  operators in semileptonic weak and electromagnetic interactions with
  nuclei}},}\ }\href {\doibase 10.1016/0092-640X(79)90003-2} {\bibfield
  {journal} {\bibinfo  {journal} {Atom. Data Nucl. Data Tabl.}\ }\textbf
  {\bibinfo {volume} {23}},\ \bibinfo {pages} {103--176} (\bibinfo {year}
  {1979})}\BibitemShut {NoStop}%
\bibitem [{\citenamefont {Edmonds}(1957)}]{edmonds1957angular}%
  \BibitemOpen
  \bibfield  {author} {\bibinfo {author} {\bibfnamefont {A.R.}\ \bibnamefont
  {Edmonds}},\ }\href {https://books.google.com/books?id=1swNAQAAIAAJ} {\emph
  {\bibinfo {title} {Angular Momentum in Quantum Mechanics}}},\ Investigations
  in physics\ (\bibinfo  {publisher} {Princeton University Press},\ \bibinfo
  {year} {1957})\BibitemShut {NoStop}%
\bibitem [{\citenamefont {Johnson}\ \emph {et~al.}(2013)\citenamefont
  {Johnson}, \citenamefont {Ormand},\ and\ \citenamefont {Krastev}}]{BIGSTICK}%
  \BibitemOpen
  \bibfield  {author} {\bibinfo {author} {\bibfnamefont {Calvin~W.}\
  \bibnamefont {Johnson}}, \bibinfo {author} {\bibfnamefont {W.~Erich}\
  \bibnamefont {Ormand}}, \ and\ \bibinfo {author} {\bibfnamefont {Plamen~G.}\
  \bibnamefont {Krastev}},\ }\bibfield  {title} {\enquote {\bibinfo {title}
  {Factorization in large-scale many-body calculations},}\ }\href@noop {}
  {\bibfield  {journal} {\bibinfo  {journal} {Computer Physics Communications}\
  }\textbf {\bibinfo {volume} {184}},\ \bibinfo {pages} {2761--2774} (\bibinfo
  {year} {2013})}\BibitemShut {NoStop}%
\bibitem [{\citenamefont {Johnson}\ \emph {et~al.}(2018)\citenamefont
  {Johnson}, \citenamefont {Ormand}, \citenamefont {McElvain},\ and\
  \citenamefont {Shan}}]{Johnson:2018hrx}%
  \BibitemOpen
  \bibfield  {author} {\bibinfo {author} {\bibfnamefont {Calvin~W.}\
  \bibnamefont {Johnson}}, \bibinfo {author} {\bibfnamefont {W.~Erich}\
  \bibnamefont {Ormand}}, \bibinfo {author} {\bibfnamefont {Kenneth~S.}\
  \bibnamefont {McElvain}}, \ and\ \bibinfo {author} {\bibfnamefont
  {Hongzhang}\ \bibnamefont {Shan}},\ }\bibfield  {title} {\enquote {\bibinfo
  {title} {{BIGSTICK: A flexible configuration-interaction shell-model
  code}},}\ }\href@noop {} {\  (\bibinfo {year} {2018})},\ \Eprint
  {http://arxiv.org/abs/1801.08432} {arXiv:1801.08432 [physics.comp-ph]}
  \BibitemShut {NoStop}%
\bibitem [{\citenamefont {Caurier}\ \emph {et~al.}(2008)\citenamefont
  {Caurier}, \citenamefont {Menendez}, \citenamefont {Nowacki},\ and\
  \citenamefont {Poves}}]{Caurier:2007wq}%
  \BibitemOpen
  \bibfield  {author} {\bibinfo {author} {\bibfnamefont {E.}~\bibnamefont
  {Caurier}}, \bibinfo {author} {\bibfnamefont {J.}~\bibnamefont {Menendez}},
  \bibinfo {author} {\bibfnamefont {F.}~\bibnamefont {Nowacki}}, \ and\
  \bibinfo {author} {\bibfnamefont {A.}~\bibnamefont {Poves}},\ }\bibfield
  {title} {\enquote {\bibinfo {title} {{The Influence of pairing on the nuclear
  matrix elements of the neutrinoless beta beta decays}},}\ }\href {\doibase
  10.1103/PhysRevLett.100.052503} {\bibfield  {journal} {\bibinfo  {journal}
  {Phys. Rev. Lett.}\ }\textbf {\bibinfo {volume} {100}},\ \bibinfo {pages}
  {052503} (\bibinfo {year} {2008})},\ \Eprint {http://arxiv.org/abs/0709.2137}
  {arXiv:0709.2137 [nucl-th]} \BibitemShut {NoStop}%
\bibitem [{\citenamefont {Caurier}\ \emph {et~al.}(2010)\citenamefont
  {Caurier}, \citenamefont {Nowacki}, \citenamefont {Poves},\ and\
  \citenamefont {Sieja}}]{Caurier:2010az}%
  \BibitemOpen
  \bibfield  {author} {\bibinfo {author} {\bibfnamefont {E.}~\bibnamefont
  {Caurier}}, \bibinfo {author} {\bibfnamefont {F.}~\bibnamefont {Nowacki}},
  \bibinfo {author} {\bibfnamefont {A.}~\bibnamefont {Poves}}, \ and\ \bibinfo
  {author} {\bibfnamefont {K.}~\bibnamefont {Sieja}},\ }\bibfield  {title}
  {\enquote {\bibinfo {title} {{Collectivity in the light Xenon isotopes: A
  shell model study}},}\ }\href {\doibase 10.1103/PhysRevC.82.064304}
  {\bibfield  {journal} {\bibinfo  {journal} {Phys. Rev. C}\ }\textbf {\bibinfo
  {volume} {82}},\ \bibinfo {pages} {064304} (\bibinfo {year} {2010})},\
  \Eprint {http://arxiv.org/abs/1009.3813} {arXiv:1009.3813 [nucl-th]}
  \BibitemShut {NoStop}%
\bibitem [{\citenamefont {Brown}\ \emph {et~al.}(2005)\citenamefont {Brown},
  \citenamefont {Stone}, \citenamefont {Stone}, \citenamefont {Towner},\ and\
  \citenamefont {Hjorth-Jensen}}]{Brown:2004xk}%
  \BibitemOpen
  \bibfield  {author} {\bibinfo {author} {\bibfnamefont {B.~Alex}\ \bibnamefont
  {Brown}}, \bibinfo {author} {\bibfnamefont {N.~J.}\ \bibnamefont {Stone}},
  \bibinfo {author} {\bibfnamefont {J.~R.}\ \bibnamefont {Stone}}, \bibinfo
  {author} {\bibfnamefont {I.~S.}\ \bibnamefont {Towner}}, \ and\ \bibinfo
  {author} {\bibfnamefont {M.}~\bibnamefont {Hjorth-Jensen}},\ }\bibfield
  {title} {\enquote {\bibinfo {title} {{Magnetic moments of the 2+(1) states
  around Sn-132}},}\ }\href {\doibase 10.1103/PhysRevC.71.044317} {\bibfield
  {journal} {\bibinfo  {journal} {Phys. Rev. C}\ }\textbf {\bibinfo {volume}
  {71}},\ \bibinfo {pages} {044317} (\bibinfo {year} {2005})},\ \bibinfo {note}
  {[Erratum: Phys.Rev.C 72, 029901 (2005)]},\ \Eprint
  {http://arxiv.org/abs/nucl-th/0411099} {arXiv:nucl-th/0411099} \BibitemShut
  {NoStop}%
\bibitem [{\citenamefont {Hjorth-Jensen}\ \emph {et~al.}(1995)\citenamefont
  {Hjorth-Jensen}, \citenamefont {Kuo},\ and\ \citenamefont
  {Osnes}}]{HJORTHJENSEN1995125}%
  \BibitemOpen
  \bibfield  {author} {\bibinfo {author} {\bibfnamefont {Morten}\ \bibnamefont
  {Hjorth-Jensen}}, \bibinfo {author} {\bibfnamefont {Thomas~T.S.}\
  \bibnamefont {Kuo}}, \ and\ \bibinfo {author} {\bibfnamefont {Eivind}\
  \bibnamefont {Osnes}},\ }\bibfield  {title} {\enquote {\bibinfo {title}
  {Realistic effective interactions for nuclear systems},}\ }\href {\doibase
  https://doi.org/10.1016/0370-1573(95)00012-6} {\bibfield  {journal} {\bibinfo
   {journal} {Physics Reports}\ }\textbf {\bibinfo {volume} {261}},\ \bibinfo
  {pages} {125--270} (\bibinfo {year} {1995})}\BibitemShut {NoStop}%
\bibitem [{\citenamefont {Machleidt}(2001)}]{Machleidt:2000ge}%
  \BibitemOpen
  \bibfield  {author} {\bibinfo {author} {\bibfnamefont {R.}~\bibnamefont
  {Machleidt}},\ }\bibfield  {title} {\enquote {\bibinfo {title} {{The High
  precision, charge dependent Bonn nucleon-nucleon potential (CD-Bonn)}},}\
  }\href {\doibase 10.1103/PhysRevC.63.024001} {\bibfield  {journal} {\bibinfo
  {journal} {Phys. Rev. C}\ }\textbf {\bibinfo {volume} {63}},\ \bibinfo
  {pages} {024001} (\bibinfo {year} {2001})},\ \Eprint
  {http://arxiv.org/abs/nucl-th/0006014} {arXiv:nucl-th/0006014} \BibitemShut
  {NoStop}%
\bibitem [{\citenamefont {Drukier}\ \emph {et~al.}(1986)\citenamefont
  {Drukier}, \citenamefont {Freese},\ and\ \citenamefont
  {Spergel}}]{Drukier:1986tm}%
  \BibitemOpen
  \bibfield  {author} {\bibinfo {author} {\bibfnamefont {A.~K.}\ \bibnamefont
  {Drukier}}, \bibinfo {author} {\bibfnamefont {Katherine}\ \bibnamefont
  {Freese}}, \ and\ \bibinfo {author} {\bibfnamefont {D.~N.}\ \bibnamefont
  {Spergel}},\ }\bibfield  {title} {\enquote {\bibinfo {title} {{Detecting Cold
  Dark Matter Candidates}},}\ }\href {\doibase 10.1103/PhysRevD.33.3495}
  {\bibfield  {journal} {\bibinfo  {journal} {Phys. Rev. D}\ }\textbf {\bibinfo
  {volume} {33}},\ \bibinfo {pages} {3495--3508} (\bibinfo {year}
  {1986})}\BibitemShut {NoStop}%
\bibitem [{\citenamefont {Freese}\ \emph {et~al.}(1988)\citenamefont {Freese},
  \citenamefont {Frieman},\ and\ \citenamefont {Gould}}]{PhysRevD.37.3388}%
  \BibitemOpen
  \bibfield  {author} {\bibinfo {author} {\bibfnamefont {Katherine}\
  \bibnamefont {Freese}}, \bibinfo {author} {\bibfnamefont {Joshua}\
  \bibnamefont {Frieman}}, \ and\ \bibinfo {author} {\bibfnamefont {Andrew}\
  \bibnamefont {Gould}},\ }\bibfield  {title} {\enquote {\bibinfo {title}
  {Signal modulation in cold-dark-matter detection},}\ }\href {\doibase
  10.1103/PhysRevD.37.3388} {\bibfield  {journal} {\bibinfo  {journal} {Phys.
  Rev. D}\ }\textbf {\bibinfo {volume} {37}},\ \bibinfo {pages} {3388--3405}
  (\bibinfo {year} {1988})}\BibitemShut {NoStop}%
\bibitem [{\citenamefont {Yoshida}\ \emph {et~al.}(2018)\citenamefont
  {Yoshida}, \citenamefont {Shimizu}, \citenamefont {Togashi},\ and\
  \citenamefont {Otsuka}}]{PhysRevC.98.061301}%
  \BibitemOpen
  \bibfield  {author} {\bibinfo {author} {\bibfnamefont {Sota}\ \bibnamefont
  {Yoshida}}, \bibinfo {author} {\bibfnamefont {Noritaka}\ \bibnamefont
  {Shimizu}}, \bibinfo {author} {\bibfnamefont {Tomoaki}\ \bibnamefont
  {Togashi}}, \ and\ \bibinfo {author} {\bibfnamefont {Takaharu}\ \bibnamefont
  {Otsuka}},\ }\bibfield  {title} {\enquote {\bibinfo {title} {Uncertainty
  quantification in the nuclear shell model},}\ }\href {\doibase
  10.1103/PhysRevC.98.061301} {\bibfield  {journal} {\bibinfo  {journal} {Phys.
  Rev. C}\ }\textbf {\bibinfo {volume} {98}},\ \bibinfo {pages} {061301}
  (\bibinfo {year} {2018})}\BibitemShut {NoStop}%
\bibitem [{\citenamefont {Fox}\ \emph {et~al.}(2020)\citenamefont {Fox},
  \citenamefont {Johnson},\ and\ \citenamefont {Perez}}]{fox_usdb}%
  \BibitemOpen
  \bibfield  {author} {\bibinfo {author} {\bibfnamefont {Jordan M.~R.}\
  \bibnamefont {Fox}}, \bibinfo {author} {\bibfnamefont {Calvin~W.}\
  \bibnamefont {Johnson}}, \ and\ \bibinfo {author} {\bibfnamefont
  {Rodrigo~Navarro}\ \bibnamefont {Perez}},\ }\bibfield  {title} {\enquote
  {\bibinfo {title} {Uncertainty quantification of an empirical shell-model
  interaction using principal component analysis},}\ }\href {\doibase
  10.1103/physrevc.101.054308} {\bibfield  {journal} {\bibinfo  {journal}
  {Physical Review C}\ }\textbf {\bibinfo {volume} {101}} (\bibinfo {year}
  {2020}),\ 10.1103/physrevc.101.054308}\BibitemShut {NoStop}%
\bibitem [{\citenamefont {Fox}\ \emph {et~al.}(2022)\citenamefont {Fox},
  \citenamefont {Johnson},\ and\ \citenamefont {Perez}}]{fox2022uncertainty}%
  \BibitemOpen
  \bibfield  {author} {\bibinfo {author} {\bibfnamefont {Jordan~MR}\
  \bibnamefont {Fox}}, \bibinfo {author} {\bibfnamefont {Calvin~W}\
  \bibnamefont {Johnson}}, \ and\ \bibinfo {author} {\bibfnamefont
  {Rodrigo~Navarro}\ \bibnamefont {Perez}},\ }\bibfield  {title} {\enquote
  {\bibinfo {title} {Uncertainty quantification of transition operators in the
  empirical shell model},}\ }\href@noop {} {\bibfield  {journal} {\bibinfo
  {journal} {arXiv preprint arXiv:2206.14956}\ } (\bibinfo {year}
  {2022})}\BibitemShut {NoStop}%
\bibitem [{\citenamefont {K{\"o}nig}\ \emph {et~al.}(2020)\citenamefont
  {K{\"o}nig}, \citenamefont {Ekstr{\"o}m}, \citenamefont {Hebeler},
  \citenamefont {Lee},\ and\ \citenamefont {Schwenk}}]{konig_ec}%
  \BibitemOpen
  \bibfield  {author} {\bibinfo {author} {\bibfnamefont {S}~\bibnamefont
  {K{\"o}nig}}, \bibinfo {author} {\bibfnamefont {A}~\bibnamefont
  {Ekstr{\"o}m}}, \bibinfo {author} {\bibfnamefont {K}~\bibnamefont {Hebeler}},
  \bibinfo {author} {\bibfnamefont {D}~\bibnamefont {Lee}}, \ and\ \bibinfo
  {author} {\bibfnamefont {A}~\bibnamefont {Schwenk}},\ }\bibfield  {title}
  {\enquote {\bibinfo {title} {Eigenvector continuation as an efficient and
  accurate emulator for uncertainty quantification},}\ }\href@noop {}
  {\bibfield  {journal} {\bibinfo  {journal} {Physics Letters B}\ }\textbf
  {\bibinfo {volume} {810}},\ \bibinfo {pages} {135814} (\bibinfo {year}
  {2020})}\BibitemShut {NoStop}%
\bibitem [{\citenamefont {Aprile}\ \emph {et~al.}(2016)\citenamefont {Aprile}
  \emph {et~al.}}]{XENON:2015gkh}%
  \BibitemOpen
  \bibfield  {author} {\bibinfo {author} {\bibfnamefont {E.}~\bibnamefont
  {Aprile}} \emph {et~al.} (\bibinfo {collaboration} {XENON}),\ }\bibfield
  {title} {\enquote {\bibinfo {title} {{Physics reach of the XENON1T dark
  matter experiment}},}\ }\href {\doibase 10.1088/1475-7516/2016/04/027}
  {\bibfield  {journal} {\bibinfo  {journal} {JCAP}\ }\textbf {\bibinfo
  {volume} {04}},\ \bibinfo {pages} {027} (\bibinfo {year} {2016})},\ \Eprint
  {http://arxiv.org/abs/1512.07501} {arXiv:1512.07501 [physics.ins-det]}
  \BibitemShut {NoStop}%
\bibitem [{\citenamefont {Aprile}\ \emph {et~al.}(2006)\citenamefont {Aprile},
  \citenamefont {Dahl}, \citenamefont {DeViveiros}, \citenamefont {Gaitskell},
  \citenamefont {Giboni}, \citenamefont {Kwong}, \citenamefont {Majewski},
  \citenamefont {Ni}, \citenamefont {Shutt},\ and\ \citenamefont
  {Yamashita}}]{Aprile:2006kx}%
  \BibitemOpen
  \bibfield  {author} {\bibinfo {author} {\bibfnamefont {E.}~\bibnamefont
  {Aprile}}, \bibinfo {author} {\bibfnamefont {C.~E.}\ \bibnamefont {Dahl}},
  \bibinfo {author} {\bibfnamefont {L.}~\bibnamefont {DeViveiros}}, \bibinfo
  {author} {\bibfnamefont {R.}~\bibnamefont {Gaitskell}}, \bibinfo {author}
  {\bibfnamefont {K.~L.}\ \bibnamefont {Giboni}}, \bibinfo {author}
  {\bibfnamefont {J.}~\bibnamefont {Kwong}}, \bibinfo {author} {\bibfnamefont
  {P.}~\bibnamefont {Majewski}}, \bibinfo {author} {\bibfnamefont {Kaixuan}\
  \bibnamefont {Ni}}, \bibinfo {author} {\bibfnamefont {T.}~\bibnamefont
  {Shutt}}, \ and\ \bibinfo {author} {\bibfnamefont {M.}~\bibnamefont
  {Yamashita}},\ }\bibfield  {title} {\enquote {\bibinfo {title} {{Simultaneous
  measurement of ionization and scintillation from nuclear recoils in liquid
  xenon as target for a dark matter experiment}},}\ }\href {\doibase
  10.1103/PhysRevLett.97.081302} {\bibfield  {journal} {\bibinfo  {journal}
  {Phys. Rev. Lett.}\ }\textbf {\bibinfo {volume} {97}},\ \bibinfo {pages}
  {081302} (\bibinfo {year} {2006})},\ \Eprint
  {http://arxiv.org/abs/astro-ph/0601552} {arXiv:astro-ph/0601552} \BibitemShut
  {NoStop}%
\bibitem [{\citenamefont {Feldman}\ and\ \citenamefont
  {Cousins}(1998)}]{Feldman:1997qc}%
  \BibitemOpen
  \bibfield  {author} {\bibinfo {author} {\bibfnamefont {Gary~J.}\ \bibnamefont
  {Feldman}}\ and\ \bibinfo {author} {\bibfnamefont {Robert~D.}\ \bibnamefont
  {Cousins}},\ }\bibfield  {title} {\enquote {\bibinfo {title} {{A Unified
  approach to the classical statistical analysis of small signals}},}\ }\href
  {\doibase 10.1103/PhysRevD.57.3873} {\bibfield  {journal} {\bibinfo
  {journal} {Phys. Rev. D}\ }\textbf {\bibinfo {volume} {57}},\ \bibinfo
  {pages} {3873--3889} (\bibinfo {year} {1998})},\ \Eprint
  {http://arxiv.org/abs/physics/9711021} {arXiv:physics/9711021} \BibitemShut
  {NoStop}%
\bibitem [{\citenamefont {Hoferichter}\ \emph {et~al.}(2016)\citenamefont
  {Hoferichter}, \citenamefont {Klos}, \citenamefont {Men\'endez},\ and\
  \citenamefont {Schwenk}}]{PhysRevD.94.063505}%
  \BibitemOpen
  \bibfield  {author} {\bibinfo {author} {\bibfnamefont {Martin}\ \bibnamefont
  {Hoferichter}}, \bibinfo {author} {\bibfnamefont {Philipp}\ \bibnamefont
  {Klos}}, \bibinfo {author} {\bibfnamefont {Javier}\ \bibnamefont
  {Men\'endez}}, \ and\ \bibinfo {author} {\bibfnamefont {Achim}\ \bibnamefont
  {Schwenk}},\ }\bibfield  {title} {\enquote {\bibinfo {title} {Analysis
  strategies for general spin-independent wimp-nucleus scattering},}\ }\href
  {\doibase 10.1103/PhysRevD.94.063505} {\bibfield  {journal} {\bibinfo
  {journal} {Phys. Rev. D}\ }\textbf {\bibinfo {volume} {94}},\ \bibinfo
  {pages} {063505} (\bibinfo {year} {2016})}\BibitemShut {NoStop}%
\bibitem [{\citenamefont {Bishara}\ \emph
  {et~al.}(2017{\natexlab{b}})\citenamefont {Bishara}, \citenamefont {Brod},
  \citenamefont {Grinstein},\ and\ \citenamefont {Zupan}}]{bishara2017quarks}%
  \BibitemOpen
  \bibfield  {author} {\bibinfo {author} {\bibfnamefont {Fady}\ \bibnamefont
  {Bishara}}, \bibinfo {author} {\bibfnamefont {Joachim}\ \bibnamefont {Brod}},
  \bibinfo {author} {\bibfnamefont {Benjamin}\ \bibnamefont {Grinstein}}, \
  and\ \bibinfo {author} {\bibfnamefont {Jure}\ \bibnamefont {Zupan}},\
  }\bibfield  {title} {\enquote {\bibinfo {title} {From quarks to nucleons in
  dark matter direct detection},}\ }\href@noop {} {\bibfield  {journal}
  {\bibinfo  {journal} {Journal of High Energy Physics}\ }\textbf {\bibinfo
  {volume} {2017}},\ \bibinfo {pages} {1--41} (\bibinfo {year}
  {2017}{\natexlab{b}})}\BibitemShut {NoStop}%
\bibitem [{\citenamefont {Xia}\ \emph {et~al.}(2019)\citenamefont {Xia},
  \citenamefont {Abdukerim}, \citenamefont {Chen}, \citenamefont {Chen},
  \citenamefont {Chen}, \citenamefont {Cui}, \citenamefont {Fang},
  \citenamefont {Fu}, \citenamefont {Giboni}, \citenamefont {Giuliani} \emph
  {et~al.}}]{xia2019pandax}%
  \BibitemOpen
  \bibfield  {author} {\bibinfo {author} {\bibfnamefont {Jingkai}\ \bibnamefont
  {Xia}}, \bibinfo {author} {\bibfnamefont {Abdusalam}\ \bibnamefont
  {Abdukerim}}, \bibinfo {author} {\bibfnamefont {Wei}\ \bibnamefont {Chen}},
  \bibinfo {author} {\bibfnamefont {Xun}\ \bibnamefont {Chen}}, \bibinfo
  {author} {\bibfnamefont {Yunhua}\ \bibnamefont {Chen}}, \bibinfo {author}
  {\bibfnamefont {Xiangyi}\ \bibnamefont {Cui}}, \bibinfo {author}
  {\bibfnamefont {Deqing}\ \bibnamefont {Fang}}, \bibinfo {author}
  {\bibfnamefont {Changbo}\ \bibnamefont {Fu}}, \bibinfo {author}
  {\bibfnamefont {Karl}\ \bibnamefont {Giboni}}, \bibinfo {author}
  {\bibfnamefont {Franco}\ \bibnamefont {Giuliani}},  \emph {et~al.},\
  }\bibfield  {title} {\enquote {\bibinfo {title} {Pandax-ii constraints on
  spin-dependent wimp-nucleon effective interactions},}\ }\href@noop {}
  {\bibfield  {journal} {\bibinfo  {journal} {Physics Letters B}\ }\textbf
  {\bibinfo {volume} {792}},\ \bibinfo {pages} {193--198} (\bibinfo {year}
  {2019})}\BibitemShut {NoStop}%
\bibitem [{\citenamefont {Alanne}\ \emph {et~al.}(2022)\citenamefont {Alanne},
  \citenamefont {Bishara}, \citenamefont {Fiaschi}, \citenamefont {Fischer},
  \citenamefont {Gorbahn},\ and\ \citenamefont {Moldanazarova}}]{alanne2022z}%
  \BibitemOpen
  \bibfield  {author} {\bibinfo {author} {\bibfnamefont {T}~\bibnamefont
  {Alanne}}, \bibinfo {author} {\bibfnamefont {F}~\bibnamefont {Bishara}},
  \bibinfo {author} {\bibfnamefont {J}~\bibnamefont {Fiaschi}}, \bibinfo
  {author} {\bibfnamefont {O}~\bibnamefont {Fischer}}, \bibinfo {author}
  {\bibfnamefont {M}~\bibnamefont {Gorbahn}}, \ and\ \bibinfo {author}
  {\bibfnamefont {U}~\bibnamefont {Moldanazarova}},\ }\bibfield  {title}
  {\enquote {\bibinfo {title} {Z'-mediated majorana dark matter: suppressed
  direct-detection rate and complementarity of lhc searches},}\ }\href@noop {}
  {\bibfield  {journal} {\bibinfo  {journal} {Journal of High Energy Physics}\
  }\textbf {\bibinfo {volume} {2022}},\ \bibinfo {pages} {1--20} (\bibinfo
  {year} {2022})}\BibitemShut {NoStop}%
\bibitem [{\citenamefont {Cheek}\ \emph {et~al.}(2023)\citenamefont {Cheek},
  \citenamefont {Price},\ and\ \citenamefont
  {Sandford}}]{cheek2023isospinviolating}%
  \BibitemOpen
  \bibfield  {author} {\bibinfo {author} {\bibfnamefont {Andrew}\ \bibnamefont
  {Cheek}}, \bibinfo {author} {\bibfnamefont {Darren~D.}\ \bibnamefont
  {Price}}, \ and\ \bibinfo {author} {\bibfnamefont {Ellen~M.}\ \bibnamefont
  {Sandford}},\ }\href@noop {} {\enquote {\bibinfo {title} {Isospin-violating
  dark matter at liquid noble detectors: new constraints, future projections,
  and an exploration of target complementarity},}\ } (\bibinfo {year} {2023}),\
  \Eprint {http://arxiv.org/abs/2302.05458} {arXiv:2302.05458 [hep-ph]}
  \BibitemShut {NoStop}%
\bibitem [{\citenamefont {Fowlie}(2019)}]{Fowlie:2018svr}%
  \BibitemOpen
  \bibfield  {author} {\bibinfo {author} {\bibfnamefont {Andrew}\ \bibnamefont
  {Fowlie}},\ }\bibfield  {title} {\enquote {\bibinfo {title} {{Non-parametric
  uncertainties in the dark matter velocity distribution}},}\ }\href {\doibase
  10.1088/1475-7516/2019/01/006} {\bibfield  {journal} {\bibinfo  {journal}
  {JCAP}\ }\textbf {\bibinfo {volume} {01}},\ \bibinfo {pages} {006} (\bibinfo
  {year} {2019})},\ \Eprint {http://arxiv.org/abs/1809.02323} {arXiv:1809.02323
  [hep-ph]} \BibitemShut {NoStop}%
\bibitem [{\citenamefont {Elekes}\ and\ \citenamefont
  {Timar}(2015)}]{ELEKES2015191}%
  \BibitemOpen
  \bibfield  {author} {\bibinfo {author} {\bibfnamefont {Zoltan}\ \bibnamefont
  {Elekes}}\ and\ \bibinfo {author} {\bibfnamefont {Janos}\ \bibnamefont
  {Timar}},\ }\bibfield  {title} {\enquote {\bibinfo {title} {Nuclear {D}ata
  {S}heets for {A} = 128},}\ }\href {\doibase
  https://doi.org/10.1016/j.nds.2015.09.002} {\bibfield  {journal} {\bibinfo
  {journal} {Nuclear Data Sheets}\ }\textbf {\bibinfo {volume} {129}},\
  \bibinfo {pages} {191--436} (\bibinfo {year} {2015})}\BibitemShut {NoStop}%
\bibitem [{\citenamefont {Timar}\ \emph {et~al.}(2014)\citenamefont {Timar},
  \citenamefont {Elekes},\ and\ \citenamefont {Singh}}]{TIMAR2014143}%
  \BibitemOpen
  \bibfield  {author} {\bibinfo {author} {\bibfnamefont {Janos}\ \bibnamefont
  {Timar}}, \bibinfo {author} {\bibfnamefont {Zoltan}\ \bibnamefont {Elekes}},
  \ and\ \bibinfo {author} {\bibfnamefont {Balraj}\ \bibnamefont {Singh}},\
  }\bibfield  {title} {\enquote {\bibinfo {title} {Nuclear {D}ata {S}heets for
  {A} = 129},}\ }\href {\doibase https://doi.org/10.1016/j.nds.2014.09.002}
  {\bibfield  {journal} {\bibinfo  {journal} {Nuclear Data Sheets}\ }\textbf
  {\bibinfo {volume} {121}},\ \bibinfo {pages} {143--394} (\bibinfo {year}
  {2014})}\BibitemShut {NoStop}%
\bibitem [{\citenamefont {Singh}(2001)}]{SINGH200133}%
  \BibitemOpen
  \bibfield  {author} {\bibinfo {author} {\bibfnamefont {Balraj}\ \bibnamefont
  {Singh}},\ }\bibfield  {title} {\enquote {\bibinfo {title} {Nuclear {D}ata
  {S}heets for {A} = 130},}\ }\href {\doibase
  https://doi.org/10.1006/ndsh.2001.0012} {\bibfield  {journal} {\bibinfo
  {journal} {Nuclear Data Sheets}\ }\textbf {\bibinfo {volume} {93}},\ \bibinfo
  {pages} {33--242} (\bibinfo {year} {2001})}\BibitemShut {NoStop}%
\bibitem [{\citenamefont {Khazov}\ \emph {et~al.}(2006)\citenamefont {Khazov},
  \citenamefont {Mitropolsky},\ and\ \citenamefont
  {Rodionov}}]{KHAZOV20062715}%
  \BibitemOpen
  \bibfield  {author} {\bibinfo {author} {\bibfnamefont {Yu.}\ \bibnamefont
  {Khazov}}, \bibinfo {author} {\bibfnamefont {I.}~\bibnamefont {Mitropolsky}},
  \ and\ \bibinfo {author} {\bibfnamefont {A.}~\bibnamefont {Rodionov}},\
  }\bibfield  {title} {\enquote {\bibinfo {title} {Nuclear {D}ata {S}heets for
  {A} = 131},}\ }\href {\doibase https://doi.org/10.1016/j.nds.2006.10.001}
  {\bibfield  {journal} {\bibinfo  {journal} {Nuclear Data Sheets}\ }\textbf
  {\bibinfo {volume} {107}},\ \bibinfo {pages} {2715--2930} (\bibinfo {year}
  {2006})}\BibitemShut {NoStop}%
\bibitem [{\citenamefont {Yu}\ \emph {et~al.}(2005)\citenamefont {Yu},
  \citenamefont {Rodionov}, \citenamefont {Sakharov},\ and\ \citenamefont
  {Singh}}]{yu2005nuclear}%
  \BibitemOpen
  \bibfield  {author} {\bibinfo {author} {\bibfnamefont {Khazov}\ \bibnamefont
  {Yu}}, \bibinfo {author} {\bibfnamefont {AA}~\bibnamefont {Rodionov}},
  \bibinfo {author} {\bibfnamefont {S}~\bibnamefont {Sakharov}}, \ and\
  \bibinfo {author} {\bibfnamefont {B}~\bibnamefont {Singh}},\ }\bibfield
  {title} {\enquote {\bibinfo {title} {Nuclear {D}ata {S}heets for {A}= 132.}}\
  }\href@noop {} {\bibfield  {journal} {\bibinfo  {journal} {Nuclear Data
  Sheets}\ }\textbf {\bibinfo {volume} {104}},\ \bibinfo {pages} {497--790}
  (\bibinfo {year} {2005})}\BibitemShut {NoStop}%
\bibitem [{\citenamefont {Sonzogni}(2004)}]{SONZOGNI20041}%
  \BibitemOpen
  \bibfield  {author} {\bibinfo {author} {\bibfnamefont {A.A.}\ \bibnamefont
  {Sonzogni}},\ }\bibfield  {title} {\enquote {\bibinfo {title} {Nuclear {D}ata
  sheets for {A} = 134},}\ }\href {\doibase
  https://doi.org/10.1016/j.nds.2004.11.001} {\bibfield  {journal} {\bibinfo
  {journal} {Nuclear Data Sheets}\ }\textbf {\bibinfo {volume} {103}},\
  \bibinfo {pages} {1--182} (\bibinfo {year} {2004})}\BibitemShut {NoStop}%
\bibitem [{\citenamefont {Mccutchan}(2018)}]{MCCUTCHAN2018331}%
  \BibitemOpen
  \bibfield  {author} {\bibinfo {author} {\bibfnamefont {E.A.}\ \bibnamefont
  {Mccutchan}},\ }\bibfield  {title} {\enquote {\bibinfo {title} {Nuclear
  {D}ata {S}heets for {A}=136},}\ }\href {\doibase
  https://doi.org/10.1016/j.nds.2018.10.002} {\bibfield  {journal} {\bibinfo
  {journal} {Nuclear Data Sheets}\ }\textbf {\bibinfo {volume} {152}},\
  \bibinfo {pages} {331--667} (\bibinfo {year} {2018})}\BibitemShut {NoStop}%
\bibitem [{\citenamefont {Brown}\ and\ \citenamefont
  {Richter}(2006)}]{PhysRevC.74.034315}%
  \BibitemOpen
  \bibfield  {author} {\bibinfo {author} {\bibfnamefont {B.~Alex}\ \bibnamefont
  {Brown}}\ and\ \bibinfo {author} {\bibfnamefont {W.~A.}\ \bibnamefont
  {Richter}},\ }\bibfield  {title} {\enquote {\bibinfo {title} {New ``{USD}''
  hamiltonians for the $\mathit{sd}$ shell},}\ }\href {\doibase
  10.1103/PhysRevC.74.034315} {\bibfield  {journal} {\bibinfo  {journal} {Phys.
  Rev. C}\ }\textbf {\bibinfo {volume} {74}},\ \bibinfo {pages} {034315}
  (\bibinfo {year} {2006})}\BibitemShut {NoStop}%
\bibitem [{\citenamefont {Hicks}\ \emph {et~al.}(2022)\citenamefont {Hicks},
  \citenamefont {Stuchbery}, \citenamefont {Churchill}, \citenamefont
  {Bandyopadhyay}, \citenamefont {Champine}, \citenamefont {Coombes},
  \citenamefont {Davoren}, \citenamefont {Ellis}, \citenamefont {Faulkner},
  \citenamefont {Lesher}, \citenamefont {Mueller}, \citenamefont
  {Mukhopadhyay}, \citenamefont {Orce}, \citenamefont {Skubis}, \citenamefont
  {Vanhoy},\ and\ \citenamefont {Yates}}]{PhysRevC.105.024329}%
  \BibitemOpen
  \bibfield  {author} {\bibinfo {author} {\bibfnamefont {S.~F.}\ \bibnamefont
  {Hicks}}, \bibinfo {author} {\bibfnamefont {A.~E.}\ \bibnamefont
  {Stuchbery}}, \bibinfo {author} {\bibfnamefont {T.~H.}\ \bibnamefont
  {Churchill}}, \bibinfo {author} {\bibfnamefont {D.}~\bibnamefont
  {Bandyopadhyay}}, \bibinfo {author} {\bibfnamefont {B.~R.}\ \bibnamefont
  {Champine}}, \bibinfo {author} {\bibfnamefont {B.~J.}\ \bibnamefont
  {Coombes}}, \bibinfo {author} {\bibfnamefont {C.~M.}\ \bibnamefont
  {Davoren}}, \bibinfo {author} {\bibfnamefont {J.~C.}\ \bibnamefont {Ellis}},
  \bibinfo {author} {\bibfnamefont {W.~M.}\ \bibnamefont {Faulkner}}, \bibinfo
  {author} {\bibfnamefont {S.~R.}\ \bibnamefont {Lesher}}, \bibinfo {author}
  {\bibfnamefont {J.~M.}\ \bibnamefont {Mueller}}, \bibinfo {author}
  {\bibfnamefont {S.}~\bibnamefont {Mukhopadhyay}}, \bibinfo {author}
  {\bibfnamefont {J.~N.}\ \bibnamefont {Orce}}, \bibinfo {author}
  {\bibfnamefont {M.~D.}\ \bibnamefont {Skubis}}, \bibinfo {author}
  {\bibfnamefont {J.~R.}\ \bibnamefont {Vanhoy}}, \ and\ \bibinfo {author}
  {\bibfnamefont {S.~W.}\ \bibnamefont {Yates}},\ }\bibfield  {title} {\enquote
  {\bibinfo {title} {Nuclear structure of $^{130}\mathrm{Te}$ from inelastic
  neutron scattering and shell model analysis},}\ }\href {\doibase
  10.1103/PhysRevC.105.024329} {\bibfield  {journal} {\bibinfo  {journal}
  {Phys. Rev. C}\ }\textbf {\bibinfo {volume} {105}},\ \bibinfo {pages}
  {024329} (\bibinfo {year} {2022})}\BibitemShut {NoStop}%
\bibitem [{\citenamefont {Rebeiro}\ \emph {et~al.}(2021)\citenamefont
  {Rebeiro}, \citenamefont {Triambak}, \citenamefont {Garrett}, \citenamefont
  {Brown}, \citenamefont {Ball}, \citenamefont {Lindsay}, \citenamefont
  {Adsley}, \citenamefont {Bildstein}, \citenamefont {Burbadge}, \citenamefont
  {Diaz-Varela}, \citenamefont {Faestermann}, \citenamefont {Hertenberger},
  \citenamefont {Jigmeddorj}, \citenamefont {Kamil}, \citenamefont {Leach},
  \citenamefont {Mabika}, \citenamefont {Ondze}, \citenamefont {Orce},
  \citenamefont {Radich},\ and\ \citenamefont {Wirth}}]{PhysRevC.104.034309}%
  \BibitemOpen
  \bibfield  {author} {\bibinfo {author} {\bibfnamefont {B.~M.}\ \bibnamefont
  {Rebeiro}}, \bibinfo {author} {\bibfnamefont {S.}~\bibnamefont {Triambak}},
  \bibinfo {author} {\bibfnamefont {P.~E.}\ \bibnamefont {Garrett}}, \bibinfo
  {author} {\bibfnamefont {B.~A.}\ \bibnamefont {Brown}}, \bibinfo {author}
  {\bibfnamefont {G.~C.}\ \bibnamefont {Ball}}, \bibinfo {author}
  {\bibfnamefont {R.}~\bibnamefont {Lindsay}}, \bibinfo {author} {\bibfnamefont
  {P.}~\bibnamefont {Adsley}}, \bibinfo {author} {\bibfnamefont
  {V.}~\bibnamefont {Bildstein}}, \bibinfo {author} {\bibfnamefont
  {C.}~\bibnamefont {Burbadge}}, \bibinfo {author} {\bibfnamefont
  {A.}~\bibnamefont {Diaz-Varela}}, \bibinfo {author} {\bibfnamefont
  {T.}~\bibnamefont {Faestermann}}, \bibinfo {author} {\bibfnamefont
  {R.}~\bibnamefont {Hertenberger}}, \bibinfo {author} {\bibfnamefont
  {B.}~\bibnamefont {Jigmeddorj}}, \bibinfo {author} {\bibfnamefont
  {M.}~\bibnamefont {Kamil}}, \bibinfo {author} {\bibfnamefont {K.~G.}\
  \bibnamefont {Leach}}, \bibinfo {author} {\bibfnamefont {P.~Z.}\ \bibnamefont
  {Mabika}}, \bibinfo {author} {\bibfnamefont {J.~C.~Nzobadila}\ \bibnamefont
  {Ondze}}, \bibinfo {author} {\bibfnamefont {J.~N.}\ \bibnamefont {Orce}},
  \bibinfo {author} {\bibfnamefont {A.}~\bibnamefont {Radich}}, \ and\ \bibinfo
  {author} {\bibfnamefont {H.-F.}\ \bibnamefont {Wirth}},\ }\bibfield  {title}
  {\enquote {\bibinfo {title} {Spectroscopy of states in $^{136}\mathrm{Ba}$
  using the $^{138}\mathrm{Ba}(p,t)$ reaction},}\ }\href {\doibase
  10.1103/PhysRevC.104.034309} {\bibfield  {journal} {\bibinfo  {journal}
  {Phys. Rev. C}\ }\textbf {\bibinfo {volume} {104}},\ \bibinfo {pages}
  {034309} (\bibinfo {year} {2021})}\BibitemShut {NoStop}%
\bibitem [{\citenamefont {Kaya}\ \emph {et~al.}(2018)\citenamefont {Kaya},
  \citenamefont {Vogt}, \citenamefont {Reiter}, \citenamefont {Siciliano},
  \citenamefont {Birkenbach}, \citenamefont {Blazhev}, \citenamefont
  {Coraggio}, \citenamefont {Teruya}, \citenamefont {Yoshinaga}, \citenamefont
  {Higashiyama}, \citenamefont {Arnswald}, \citenamefont {Bazzacco},
  \citenamefont {Bracco}, \citenamefont {Bruyneel}, \citenamefont {Corradi},
  \citenamefont {Crespi}, \citenamefont {de~Angelis}, \citenamefont {Eberth},
  \citenamefont {Farnea}, \citenamefont {Fioretto}, \citenamefont {Fransen},
  \citenamefont {Fu}, \citenamefont {Gadea}, \citenamefont {Gargano},
  \citenamefont {Giaz}, \citenamefont {G\"orgen}, \citenamefont {Gottardo},
  \citenamefont {Hady\ifmmode \acute{n}\else
  \'{n}\fi{}ska-Kl\ifmmode~\mbox{\k{e}}\else \k{e}\fi{}k}, \citenamefont
  {Hess}, \citenamefont {Hetzenegger}, \citenamefont {Hirsch}, \citenamefont
  {Itaco}, \citenamefont {John}, \citenamefont {Jolie}, \citenamefont
  {Jungclaus}, \citenamefont {Korten}, \citenamefont {Leoni}, \citenamefont
  {Lewandowski}, \citenamefont {Lunardi}, \citenamefont {Menegazzo},
  \citenamefont {Mengoni}, \citenamefont {Michelagnoli}, \citenamefont
  {Mijatovi\ifmmode~\acute{c}\else \'{c}\fi{}}, \citenamefont {Montagnoli},
  \citenamefont {Montanari}, \citenamefont {M\"uller-Gatermann}, \citenamefont
  {Napoli}, \citenamefont {Podoly\'ak}, \citenamefont {Pollarolo},
  \citenamefont {Pullia}, \citenamefont {Queiser}, \citenamefont {Recchia},
  \citenamefont {Rosiak}, \citenamefont {Saed-Samii}, \citenamefont
  {\ifmmode~\mbox{\c{S}}\else \c{S}\fi{}ahin}, \citenamefont {Scarlassara},
  \citenamefont {Schneiders}, \citenamefont {Seidlitz}, \citenamefont
  {Siebeck}, \citenamefont {Smith}, \citenamefont {S\"oderstr\"om},
  \citenamefont {Stefanini}, \citenamefont {Steinbach}, \citenamefont
  {Stezowski}, \citenamefont {Szilner}, \citenamefont {Szpak}, \citenamefont
  {Ur}, \citenamefont {Valiente-Dob\'on}, \citenamefont {Wolf},\ and\
  \citenamefont {Zell}}]{PhysRevC.98.014309}%
  \BibitemOpen
  \bibfield  {author} {\bibinfo {author} {\bibfnamefont {L.}~\bibnamefont
  {Kaya}}, \bibinfo {author} {\bibfnamefont {A.}~\bibnamefont {Vogt}}, \bibinfo
  {author} {\bibfnamefont {P.}~\bibnamefont {Reiter}}, \bibinfo {author}
  {\bibfnamefont {M.}~\bibnamefont {Siciliano}}, \bibinfo {author}
  {\bibfnamefont {B.}~\bibnamefont {Birkenbach}}, \bibinfo {author}
  {\bibfnamefont {A.}~\bibnamefont {Blazhev}}, \bibinfo {author} {\bibfnamefont
  {L.}~\bibnamefont {Coraggio}}, \bibinfo {author} {\bibfnamefont
  {E.}~\bibnamefont {Teruya}}, \bibinfo {author} {\bibfnamefont
  {N.}~\bibnamefont {Yoshinaga}}, \bibinfo {author} {\bibfnamefont
  {K.}~\bibnamefont {Higashiyama}}, \bibinfo {author} {\bibfnamefont
  {K.}~\bibnamefont {Arnswald}}, \bibinfo {author} {\bibfnamefont
  {D.}~\bibnamefont {Bazzacco}}, \bibinfo {author} {\bibfnamefont
  {A.}~\bibnamefont {Bracco}}, \bibinfo {author} {\bibfnamefont
  {B.}~\bibnamefont {Bruyneel}}, \bibinfo {author} {\bibfnamefont
  {L.}~\bibnamefont {Corradi}}, \bibinfo {author} {\bibfnamefont {F.~C.~L.}\
  \bibnamefont {Crespi}}, \bibinfo {author} {\bibfnamefont {G.}~\bibnamefont
  {de~Angelis}}, \bibinfo {author} {\bibfnamefont {J.}~\bibnamefont {Eberth}},
  \bibinfo {author} {\bibfnamefont {E.}~\bibnamefont {Farnea}}, \bibinfo
  {author} {\bibfnamefont {E.}~\bibnamefont {Fioretto}}, \bibinfo {author}
  {\bibfnamefont {C.}~\bibnamefont {Fransen}}, \bibinfo {author} {\bibfnamefont
  {B.}~\bibnamefont {Fu}}, \bibinfo {author} {\bibfnamefont {A.}~\bibnamefont
  {Gadea}}, \bibinfo {author} {\bibfnamefont {A.}~\bibnamefont {Gargano}},
  \bibinfo {author} {\bibfnamefont {A.}~\bibnamefont {Giaz}}, \bibinfo {author}
  {\bibfnamefont {A.}~\bibnamefont {G\"orgen}}, \bibinfo {author}
  {\bibfnamefont {A.}~\bibnamefont {Gottardo}}, \bibinfo {author}
  {\bibfnamefont {K.}~\bibnamefont {Hady\ifmmode \acute{n}\else
  \'{n}\fi{}ska-Kl\ifmmode~\mbox{\k{e}}\else \k{e}\fi{}k}}, \bibinfo {author}
  {\bibfnamefont {H.}~\bibnamefont {Hess}}, \bibinfo {author} {\bibfnamefont
  {R.}~\bibnamefont {Hetzenegger}}, \bibinfo {author} {\bibfnamefont
  {R.}~\bibnamefont {Hirsch}}, \bibinfo {author} {\bibfnamefont
  {N.}~\bibnamefont {Itaco}}, \bibinfo {author} {\bibfnamefont {P.~R.}\
  \bibnamefont {John}}, \bibinfo {author} {\bibfnamefont {J.}~\bibnamefont
  {Jolie}}, \bibinfo {author} {\bibfnamefont {A.}~\bibnamefont {Jungclaus}},
  \bibinfo {author} {\bibfnamefont {W.}~\bibnamefont {Korten}}, \bibinfo
  {author} {\bibfnamefont {S.}~\bibnamefont {Leoni}}, \bibinfo {author}
  {\bibfnamefont {L.}~\bibnamefont {Lewandowski}}, \bibinfo {author}
  {\bibfnamefont {S.}~\bibnamefont {Lunardi}}, \bibinfo {author} {\bibfnamefont
  {R.}~\bibnamefont {Menegazzo}}, \bibinfo {author} {\bibfnamefont
  {D.}~\bibnamefont {Mengoni}}, \bibinfo {author} {\bibfnamefont
  {C.}~\bibnamefont {Michelagnoli}}, \bibinfo {author} {\bibfnamefont
  {T.}~\bibnamefont {Mijatovi\ifmmode~\acute{c}\else \'{c}\fi{}}}, \bibinfo
  {author} {\bibfnamefont {G.}~\bibnamefont {Montagnoli}}, \bibinfo {author}
  {\bibfnamefont {D.}~\bibnamefont {Montanari}}, \bibinfo {author}
  {\bibfnamefont {C.}~\bibnamefont {M\"uller-Gatermann}}, \bibinfo {author}
  {\bibfnamefont {D.}~\bibnamefont {Napoli}}, \bibinfo {author} {\bibfnamefont
  {Zs.}\ \bibnamefont {Podoly\'ak}}, \bibinfo {author} {\bibfnamefont
  {G.}~\bibnamefont {Pollarolo}}, \bibinfo {author} {\bibfnamefont
  {A.}~\bibnamefont {Pullia}}, \bibinfo {author} {\bibfnamefont
  {M.}~\bibnamefont {Queiser}}, \bibinfo {author} {\bibfnamefont
  {F.}~\bibnamefont {Recchia}}, \bibinfo {author} {\bibfnamefont
  {D.}~\bibnamefont {Rosiak}}, \bibinfo {author} {\bibfnamefont
  {N.}~\bibnamefont {Saed-Samii}}, \bibinfo {author} {\bibfnamefont
  {E.}~\bibnamefont {\ifmmode~\mbox{\c{S}}\else \c{S}\fi{}ahin}}, \bibinfo
  {author} {\bibfnamefont {F.}~\bibnamefont {Scarlassara}}, \bibinfo {author}
  {\bibfnamefont {D.}~\bibnamefont {Schneiders}}, \bibinfo {author}
  {\bibfnamefont {M.}~\bibnamefont {Seidlitz}}, \bibinfo {author}
  {\bibfnamefont {B.}~\bibnamefont {Siebeck}}, \bibinfo {author} {\bibfnamefont
  {J.~F.}\ \bibnamefont {Smith}}, \bibinfo {author} {\bibfnamefont {P.-A.}\
  \bibnamefont {S\"oderstr\"om}}, \bibinfo {author} {\bibfnamefont {A.~M.}\
  \bibnamefont {Stefanini}}, \bibinfo {author} {\bibfnamefont {T.}~\bibnamefont
  {Steinbach}}, \bibinfo {author} {\bibfnamefont {O.}~\bibnamefont
  {Stezowski}}, \bibinfo {author} {\bibfnamefont {S.}~\bibnamefont {Szilner}},
  \bibinfo {author} {\bibfnamefont {B.}~\bibnamefont {Szpak}}, \bibinfo
  {author} {\bibfnamefont {C.}~\bibnamefont {Ur}}, \bibinfo {author}
  {\bibfnamefont {J.~J.}\ \bibnamefont {Valiente-Dob\'on}}, \bibinfo {author}
  {\bibfnamefont {K.}~\bibnamefont {Wolf}}, \ and\ \bibinfo {author}
  {\bibfnamefont {K.~O.}\ \bibnamefont {Zell}},\ }\bibfield  {title} {\enquote
  {\bibinfo {title} {High-spin structure in the transitional nucleus
  $^{131}\mathrm{Xe}$: Competitive neutron and proton alignment in the vicinity
  of the ${N}=82$ shell closure},}\ }\href {\doibase
  10.1103/PhysRevC.98.014309} {\bibfield  {journal} {\bibinfo  {journal} {Phys.
  Rev. C}\ }\textbf {\bibinfo {volume} {98}},\ \bibinfo {pages} {014309}
  (\bibinfo {year} {2018})}\BibitemShut {NoStop}%
\bibitem [{\citenamefont {Kumar}\ \emph {et~al.}(2022)\citenamefont {Kumar},
  \citenamefont {Bhattacharjee}, \citenamefont {Alam}, \citenamefont {Basak},
  \citenamefont {Gerhard}, \citenamefont {Knafla}, \citenamefont {Esmaylzadeh},
  \citenamefont {Ley}, \citenamefont {Dunkel}, \citenamefont {Schomaker},
  \citenamefont {R\'egis}, \citenamefont {Jolie}, \citenamefont {Kim},
  \citenamefont {K\"oster}, \citenamefont {Simpson},\ and\ \citenamefont
  {Fraile}}]{PhysRevC.106.034306}%
  \BibitemOpen
  \bibfield  {author} {\bibinfo {author} {\bibfnamefont {D.}~\bibnamefont
  {Kumar}}, \bibinfo {author} {\bibfnamefont {T.}~\bibnamefont
  {Bhattacharjee}}, \bibinfo {author} {\bibfnamefont {S.~S.}\ \bibnamefont
  {Alam}}, \bibinfo {author} {\bibfnamefont {S.}~\bibnamefont {Basak}},
  \bibinfo {author} {\bibfnamefont {L.}~\bibnamefont {Gerhard}}, \bibinfo
  {author} {\bibfnamefont {L.}~\bibnamefont {Knafla}}, \bibinfo {author}
  {\bibfnamefont {A.}~\bibnamefont {Esmaylzadeh}}, \bibinfo {author}
  {\bibfnamefont {M.}~\bibnamefont {Ley}}, \bibinfo {author} {\bibfnamefont
  {F.}~\bibnamefont {Dunkel}}, \bibinfo {author} {\bibfnamefont
  {K.}~\bibnamefont {Schomaker}}, \bibinfo {author} {\bibfnamefont {J.~M.}\
  \bibnamefont {R\'egis}}, \bibinfo {author} {\bibfnamefont {J.}~\bibnamefont
  {Jolie}}, \bibinfo {author} {\bibfnamefont {Y.~H.}\ \bibnamefont {Kim}},
  \bibinfo {author} {\bibfnamefont {U.}~\bibnamefont {K\"oster}}, \bibinfo
  {author} {\bibfnamefont {G.~S.}\ \bibnamefont {Simpson}}, \ and\ \bibinfo
  {author} {\bibfnamefont {L.~M.}\ \bibnamefont {Fraile}},\ }\bibfield  {title}
  {\enquote {\bibinfo {title} {Lifetimes and transition probabilities for
  low-lying yrast levels in $^{130,132}\mathrm{Te}$},}\ }\href {\doibase
  10.1103/PhysRevC.106.034306} {\bibfield  {journal} {\bibinfo  {journal}
  {Phys. Rev. C}\ }\textbf {\bibinfo {volume} {106}},\ \bibinfo {pages}
  {034306} (\bibinfo {year} {2022})}\BibitemShut {NoStop}%
\bibitem [{\citenamefont {Prill}\ \emph {et~al.}(2022)\citenamefont {Prill},
  \citenamefont {Bohn}, \citenamefont {Everwyn}, \citenamefont {H\"afner},
  \citenamefont {Heim}, \citenamefont {Spieker}, \citenamefont {Weinert},
  \citenamefont {Wilhelmy},\ and\ \citenamefont
  {Zilges}}]{PhysRevC.105.034319}%
  \BibitemOpen
  \bibfield  {author} {\bibinfo {author} {\bibfnamefont {Sarah}\ \bibnamefont
  {Prill}}, \bibinfo {author} {\bibfnamefont {Anna}\ \bibnamefont {Bohn}},
  \bibinfo {author} {\bibfnamefont {Vera}\ \bibnamefont {Everwyn}}, \bibinfo
  {author} {\bibfnamefont {Guillaume}\ \bibnamefont {H\"afner}}, \bibinfo
  {author} {\bibfnamefont {Felix}\ \bibnamefont {Heim}}, \bibinfo {author}
  {\bibfnamefont {Mark}\ \bibnamefont {Spieker}}, \bibinfo {author}
  {\bibfnamefont {Michael}\ \bibnamefont {Weinert}}, \bibinfo {author}
  {\bibfnamefont {Julius}\ \bibnamefont {Wilhelmy}}, \ and\ \bibinfo {author}
  {\bibfnamefont {Andreas}\ \bibnamefont {Zilges}},\ }\bibfield  {title}
  {\enquote {\bibinfo {title} {Lifetime analysis of $^{128,130}\mathrm{Te}$ via
  the doppler-shift attenuation method},}\ }\href {\doibase
  10.1103/PhysRevC.105.034319} {\bibfield  {journal} {\bibinfo  {journal}
  {Phys. Rev. C}\ }\textbf {\bibinfo {volume} {105}},\ \bibinfo {pages}
  {034319} (\bibinfo {year} {2022})}\BibitemShut {NoStop}%
\bibitem [{\citenamefont {Alam}\ \emph {et~al.}(2019)\citenamefont {Alam},
  \citenamefont {Bhattacharjee}, \citenamefont {Banerjee}, \citenamefont
  {Saha}, \citenamefont {Das}, \citenamefont {Sarkar},\ and\ \citenamefont
  {Sarkar}}]{PhysRevC.99.014306}%
  \BibitemOpen
  \bibfield  {author} {\bibinfo {author} {\bibfnamefont {S.~S.}\ \bibnamefont
  {Alam}}, \bibinfo {author} {\bibfnamefont {T.}~\bibnamefont {Bhattacharjee}},
  \bibinfo {author} {\bibfnamefont {D.}~\bibnamefont {Banerjee}}, \bibinfo
  {author} {\bibfnamefont {A.}~\bibnamefont {Saha}}, \bibinfo {author}
  {\bibfnamefont {S.}~\bibnamefont {Das}}, \bibinfo {author} {\bibfnamefont
  {M.~Saha}\ \bibnamefont {Sarkar}}, \ and\ \bibinfo {author} {\bibfnamefont
  {S.}~\bibnamefont {Sarkar}},\ }\bibfield  {title} {\enquote {\bibinfo {title}
  {Lifetimes and transition probabilities for the low-lying states in
  $^{131}\mathrm{I}$ and $^{132}\mathrm{Xe}$},}\ }\href {\doibase
  10.1103/PhysRevC.99.014306} {\bibfield  {journal} {\bibinfo  {journal} {Phys.
  Rev. C}\ }\textbf {\bibinfo {volume} {99}},\ \bibinfo {pages} {014306}
  (\bibinfo {year} {2019})}\BibitemShut {NoStop}%
\bibitem [{\citenamefont {Klos}\ \emph {et~al.}(2013)\citenamefont {Klos},
  \citenamefont {Men\'endez}, \citenamefont {Gazit},\ and\ \citenamefont
  {Schwenk}}]{PhysRevD.88.083516}%
  \BibitemOpen
  \bibfield  {author} {\bibinfo {author} {\bibfnamefont {P.}~\bibnamefont
  {Klos}}, \bibinfo {author} {\bibfnamefont {J.}~\bibnamefont {Men\'endez}},
  \bibinfo {author} {\bibfnamefont {D.}~\bibnamefont {Gazit}}, \ and\ \bibinfo
  {author} {\bibfnamefont {A.}~\bibnamefont {Schwenk}},\ }\bibfield  {title}
  {\enquote {\bibinfo {title} {Large-scale nuclear structure calculations for
  spin-dependent wimp scattering with chiral effective field theory
  currents},}\ }\href {\doibase 10.1103/PhysRevD.88.083516} {\bibfield
  {journal} {\bibinfo  {journal} {Phys. Rev. D}\ }\textbf {\bibinfo {volume}
  {88}},\ \bibinfo {pages} {083516} (\bibinfo {year} {2013})}\BibitemShut
  {NoStop}%
\bibitem [{\citenamefont {Brussard}\ and\ \citenamefont
  {Glaudemans}(1977)}]{BG77}%
  \BibitemOpen
  \bibfield  {author} {\bibinfo {author} {\bibfnamefont {P.J.}\ \bibnamefont
  {Brussard}}\ and\ \bibinfo {author} {\bibfnamefont {P.W.M.}\ \bibnamefont
  {Glaudemans}},\ }\href@noop {} {\emph {\bibinfo {title} {Shell-model
  applications in nuclear spectroscopy}}}\ (\bibinfo  {publisher}
  {North-Holland Publishing Company, Amsterdam},\ \bibinfo {year}
  {1977})\BibitemShut {NoStop}%
\bibitem [{\citenamefont {Gordon}\ \emph {et~al.}(1975)\citenamefont {Gordon},
  \citenamefont {Eytel}, \citenamefont {de~Waard},\ and\ \citenamefont
  {Murnick}}]{PhysRevC.12.628}%
  \BibitemOpen
  \bibfield  {author} {\bibinfo {author} {\bibfnamefont {D.~M.}\ \bibnamefont
  {Gordon}}, \bibinfo {author} {\bibfnamefont {L.~S.}\ \bibnamefont {Eytel}},
  \bibinfo {author} {\bibfnamefont {H.}~\bibnamefont {de~Waard}}, \ and\
  \bibinfo {author} {\bibfnamefont {D.~E.}\ \bibnamefont {Murnick}},\
  }\bibfield  {title} {\enquote {\bibinfo {title}
  {Perturbed-angular-correlation studies of ${2}^{+}$ levels of even xenon
  isotopes},}\ }\href {\doibase 10.1103/PhysRevC.12.628} {\bibfield  {journal}
  {\bibinfo  {journal} {Phys. Rev. C}\ }\textbf {\bibinfo {volume} {12}},\
  \bibinfo {pages} {628--636} (\bibinfo {year} {1975})}\BibitemShut {NoStop}%
\bibitem [{\citenamefont {Brinkmann}\ \emph {et~al.}(1962)\citenamefont
  {Brinkmann}, \citenamefont {Brun},\ and\ \citenamefont
  {Staub}}]{Brinkmann62}%
  \BibitemOpen
  \bibfield  {author} {\bibinfo {author} {\bibfnamefont {D.}~\bibnamefont
  {Brinkmann}}, \bibinfo {author} {\bibfnamefont {E.}~\bibnamefont {Brun}}, \
  and\ \bibinfo {author} {\bibfnamefont {H.~H.}\ \bibnamefont {Staub}},\
  }\bibfield  {title} {\enquote {\bibinfo {title} {Kernresonanz im
  gasf{\"o}rmigen {X}enon},}\ }\href@noop {} {\bibfield  {journal} {\bibinfo
  {journal} {Helv.~Phys.~Acta}\ }\textbf {\bibinfo {volume} {35}},\ \bibinfo
  {pages} {431--436} (\bibinfo {year} {1962})}\BibitemShut {NoStop}%
\bibitem [{\citenamefont {Van~Rossum}\ \emph {et~al.}(1974)\citenamefont
  {Van~Rossum}, \citenamefont {Langouche}, \citenamefont {Pattyn},
  \citenamefont {Dumont}, \citenamefont {Odeurs}, \citenamefont {Meykens},
  \citenamefont {Boolchand},\ and\ \citenamefont {Coussement}}]{van1974study}%
  \BibitemOpen
  \bibfield  {author} {\bibinfo {author} {\bibfnamefont {M}~\bibnamefont
  {Van~Rossum}}, \bibinfo {author} {\bibfnamefont {G}~\bibnamefont
  {Langouche}}, \bibinfo {author} {\bibfnamefont {H}~\bibnamefont {Pattyn}},
  \bibinfo {author} {\bibfnamefont {G}~\bibnamefont {Dumont}}, \bibinfo
  {author} {\bibfnamefont {J}~\bibnamefont {Odeurs}}, \bibinfo {author}
  {\bibfnamefont {A}~\bibnamefont {Meykens}}, \bibinfo {author} {\bibfnamefont
  {P}~\bibnamefont {Boolchand}}, \ and\ \bibinfo {author} {\bibfnamefont
  {R}~\bibnamefont {Coussement}},\ }\bibfield  {title} {\enquote {\bibinfo
  {title} {Study of the magnetic interaction at 129$^m${X}e implanted in
  iron},}\ }\href@noop {} {\bibfield  {journal} {\bibinfo  {journal} {Le
  Journal de Physique Colloques}\ }\textbf {\bibinfo {volume} {35}},\ \bibinfo
  {pages} {C6--301} (\bibinfo {year} {1974})}\BibitemShut {NoStop}%
\bibitem [{\citenamefont {Jakob}\ \emph {et~al.}(2002)\citenamefont {Jakob},
  \citenamefont {Benczer-Koller}, \citenamefont {Kumbartzki}, \citenamefont
  {Holden}, \citenamefont {Mertzimekis}, \citenamefont {Speidel}, \citenamefont
  {Ernst}, \citenamefont {Stuchbery}, \citenamefont {Pakou}, \citenamefont
  {Maier-Komor}, \citenamefont {Macchiavelli}, \citenamefont {McMahan},
  \citenamefont {Phair},\ and\ \citenamefont {Lee}}]{PhysRevC.65.024316}%
  \BibitemOpen
  \bibfield  {author} {\bibinfo {author} {\bibfnamefont {G.}~\bibnamefont
  {Jakob}}, \bibinfo {author} {\bibfnamefont {N.}~\bibnamefont
  {Benczer-Koller}}, \bibinfo {author} {\bibfnamefont {G.}~\bibnamefont
  {Kumbartzki}}, \bibinfo {author} {\bibfnamefont {J.}~\bibnamefont {Holden}},
  \bibinfo {author} {\bibfnamefont {T.~J.}\ \bibnamefont {Mertzimekis}},
  \bibinfo {author} {\bibfnamefont {K.-H.}\ \bibnamefont {Speidel}}, \bibinfo
  {author} {\bibfnamefont {R.}~\bibnamefont {Ernst}}, \bibinfo {author}
  {\bibfnamefont {A.~E.}\ \bibnamefont {Stuchbery}}, \bibinfo {author}
  {\bibfnamefont {A.}~\bibnamefont {Pakou}}, \bibinfo {author} {\bibfnamefont
  {P.}~\bibnamefont {Maier-Komor}}, \bibinfo {author} {\bibfnamefont
  {A.}~\bibnamefont {Macchiavelli}}, \bibinfo {author} {\bibfnamefont
  {M.}~\bibnamefont {McMahan}}, \bibinfo {author} {\bibfnamefont
  {L.}~\bibnamefont {Phair}}, \ and\ \bibinfo {author} {\bibfnamefont {I.~Y.}\
  \bibnamefont {Lee}},\ }\bibfield  {title} {\enquote {\bibinfo {title}
  {Evidence for proton excitations in ${}^{130,132,134,136}\mathrm{Xe}$
  isotopes from measurements of $g$ factors of ${2}_{1}^{+}$ and ${4}_{1}^{+}$
  states},}\ }\href {\doibase 10.1103/PhysRevC.65.024316} {\bibfield  {journal}
  {\bibinfo  {journal} {Phys. Rev. C}\ }\textbf {\bibinfo {volume} {65}},\
  \bibinfo {pages} {024316} (\bibinfo {year} {2002})}\BibitemShut {NoStop}%
\bibitem [{\citenamefont {Berant}\ \emph {et~al.}(1985)\citenamefont {Berant},
  \citenamefont {Wolf}, \citenamefont {Hill}, \citenamefont {Wohn},
  \citenamefont {Gill}, \citenamefont {Mach}, \citenamefont {Rafailovich},
  \citenamefont {Kruse}, \citenamefont {Wildenthal}, \citenamefont {Peaslee},
  \citenamefont {Aprahamian}, \citenamefont {Goulden},\ and\ \citenamefont
  {Chung}}]{PhysRevC.31.570}%
  \BibitemOpen
  \bibfield  {author} {\bibinfo {author} {\bibfnamefont {Z.}~\bibnamefont
  {Berant}}, \bibinfo {author} {\bibfnamefont {A.}~\bibnamefont {Wolf}},
  \bibinfo {author} {\bibfnamefont {John~C.}\ \bibnamefont {Hill}}, \bibinfo
  {author} {\bibfnamefont {F.~K.}\ \bibnamefont {Wohn}}, \bibinfo {author}
  {\bibfnamefont {R.~L.}\ \bibnamefont {Gill}}, \bibinfo {author}
  {\bibfnamefont {H.}~\bibnamefont {Mach}}, \bibinfo {author} {\bibfnamefont
  {M.}~\bibnamefont {Rafailovich}}, \bibinfo {author} {\bibfnamefont
  {H.}~\bibnamefont {Kruse}}, \bibinfo {author} {\bibfnamefont {B.~H.}\
  \bibnamefont {Wildenthal}}, \bibinfo {author} {\bibfnamefont
  {G.}~\bibnamefont {Peaslee}}, \bibinfo {author} {\bibfnamefont
  {A.}~\bibnamefont {Aprahamian}}, \bibinfo {author} {\bibfnamefont
  {J.}~\bibnamefont {Goulden}}, \ and\ \bibinfo {author} {\bibfnamefont
  {C.}~\bibnamefont {Chung}},\ }\bibfield  {title} {\enquote {\bibinfo {title}
  {$g$ factor of ${4}_{1}^{+}$ states in the {N}=82 isotones
  $^{136}\mathrm{Xe}$ and $^{138}\mathrm{Ba}$},}\ }\href {\doibase
  10.1103/PhysRevC.31.570} {\bibfield  {journal} {\bibinfo  {journal} {Phys.
  Rev. C}\ }\textbf {\bibinfo {volume} {31}},\ \bibinfo {pages} {570--574}
  (\bibinfo {year} {1985})}\BibitemShut {NoStop}%
\end{thebibliography}%

\end{document}